\documentclass[11pt]{article}

\usepackage{times}
\usepackage[top=25truemm,bottom=20truemm,left=20truemm,right=20truemm]{geometry}

\usepackage{graphicx}
\graphicspath{{./figs/}}  
\usepackage[usenames]{color}
\usepackage{amsmath,amssymb,amsfonts,amsthm,amscd}	
\usepackage{bm,braket,array}
\usepackage{multirow}
\usepackage[all]{xy}
\usepackage{slashed}
\usepackage[format=hang,justification=raggedright]{caption} 
\usepackage{mathabx} 

\theoremstyle{definition}

\theoremstyle{remark}

\def\im{\mbox{Im }}

\def\tr{{\rm tr\,}}

\def\p{\partial}

\def\wt{\widetilde}
\def\wh{\widehat}
\def\ker{\mbox{Ker }}

\newcommand{\g}{\gamma}
\newcommand{\s}{\sigma}
\newcommand{\G}{\Gamma}

\newcommand{\C}{\mathbb{C}}

\newcommand{\Z}{\mathbb{Z}}

\newcommand{\bk}{\bm{k}}

\def\widebar{\accentset{{\cc@style\underline{\mskip10mu}}}} 
\def\wideubar{\underaccent{{\cc@style\underline{\mskip10mu}}}} 

\title{The classification of surface states of topological insulators and superconductors with magnetic point group symmetry} 
\author{
Ken Shiozaki\thanks{ken.shiozaki@yukawa.kyoto-u.ac.jp}\\
\ \\
{\it Yukawa Institute for Theoretical Physics, Kyoto University, Kyoto 606-8502, Japan}
}
\date{\today} 
\begin{document} 
\maketitle
\begin{abstract}
We present the exhaustive classification of surface states of topological insulators and superconductors protected by crystallographic magnetic point group symmetry in three spatial dimensions. 
Recently, Cornfeld and Chapman [Phys. Rev. B {\bf 99}, 075105 (2019)] pointed out that the topological classification of mass terms of the Dirac Hamiltonian with point group symmetry is recast as the extension problem of the Clifford algebra, and we use their results extensively. 
Comparing two-types of Dirac Hamiltonians with and without the mass-hedgehog potential, we establish the irreducible character formula to read off which Hamiltonian in the whole $K$-group belongs to fourth-order topological phases, which are atomic insulators localized at the center of the point group.\end{abstract}

\setcounter{tocdepth}{2}
\tableofcontents


\section{Introduction}

Topological insulators (TIs) and topological superconductors (TSCs) are topological classes of gapped Hamiltonians of free fermions~\cite{HasanKaneRMP,QiZhangRMP}. 
Two gapped Hamiltonians are said to belong to the same topological phases if there exists a continuous path of Hamiltonians interpolating between them that preserves both symmetry and the energy gap. 
For TIs/TSCs protected solely by onsite symmetry, symmetry groups composed of transformations that keep the spatial position invariant, the classification problem is recast as the extension problem of the Clifford algebra~\cite{RyuTenFold,KitaevPeriodic,Stone2010,WenFreeFermion,MorimotoClifford}, and it is shown that a nontrivial TI/TSC exhibits a robust gapless state on its boundary. 

With crystalline symmetry, which involves a group element that changes a spatial position, the relationship between TIs/TSCs and the gapless boundary states is more involved. 
This is because atomic insulators, which are occupied states of localized atomic orbitals, contribute to the topological classification while they are irrelevant to boundary gapless states. 
The best way to think of the structure of the bulk-boundary correspondence in the presence of crystalline symmetry is the filtration $\cdots \subset K'' \subset K' \subset K$ for the classification of TIs/TSCs with respect to the space dimension on which the TI/TSC is defined~\cite{HermeleBuilding, LukaHigherOrder, KSGeneHomo}. 
Precisely, for $d$-space dimensions, let $K^{(p)}$ be the abelian group composed of TIs/TSCs that are adiabatically connected to TIs/TSCs supported on a subspace with dimension less than or equal to $(d-p)$. 
The quotient group $K^{(p)}/K^{(p+1)}$ captures TIs/TSCs supported exactly on a $(d-p)$ dimensional real-space submanifold, which are called $(p+1)$th-order TIs/TSCs~\cite{BenalcazarScience,FangFu17,SchindlerHigher}, and such phases host gapless boundary states if $p<d$. 
In particular, the group $K^{(d)}$, the $(d+1)$th-order TIs/TSCs, represents atomic insulators. 
Therefore, the quotient group $K/K^{(d)}$ provides the classification of TIs/TSCs with a gapless boundary state.
See Refs.~\cite{KhalafPRX18AII,SongMapping,HermeleTopologicalCrystal} for the classification of surface states of time-reversal (TR) symmetric TIs in spinful electrons with the connection to the symmetry-based indicator~\cite{Haruki230}. 

Recently, Cornfeld and Chapman pointed out that for arbitrary point group symmetry (namely, a subgroup of the orthogonal group keeping a crystal structure), symmetry operators can be onsite with keeping the topological classification, resulting in that the abelian group $K$ of the classification of TIs/TSCs can be computed by the extension problem of the Clifford algebra as well as the classification of TIs/TSCs with onsite symmetry~\cite{CornfeldPointGroup}. 

In this paper, for magnetic point group symmetry, we show that in addition to the group $K$ of the entire classification, the group $K^{(d)}$ of $d$th-order TIs/TSCs can also be computed in a canonical way as a subgroup of $K$. 
Taking the quotient of $K$ by $K^{(d)}$, one can get the classification of TIs/TSCs hosting a gapless boundary state. 

Throughout this paper, the classification of band structures means that in the sense of the $K$-theory.
Every classification is an abelian group, measures the formal difference between two TIs/TSCs, and is stable under adding a common TI/TSC. 
Excepted for Sec.~\ref{sec:ex}, we consider TIs/TSCs in three space dimensions. 
The same approach does work for generic space dimensions. 

The plan of this paper is as follows. 
In Sec.~\ref{sec:ex}, we illustrate how the entire group $K$ and the group $K'$ of atomic insulators are computed for a simple example in 1d. 
Sec.~\ref{sec:formulation} is devoted to establishing the irreducible character formulas to compute the groups $K$ and $K'''$ for generic magnetic point group (MPG) symmetry. 
In Sec.~\ref{sec:K}, we reformulate Cornfeld and Chapman's prescription so that it can be applied to generic MPG symmetry with arbitrary factor systems. 
In Sec.\ref{sec:ti/tscs}, we apply our formalism to insulators and superconductors in three space dimensions to get the complete classification of the gapless surface states. 
The classification tables are summarized in Tables~\ref{tab:EAZ_TI_spinless}-\ref{tab:surf_states_sc_spinful} in Appendix~\ref{app:tables}. 
We summarize this paper and suggest future directions in Sec.~\ref{sec:conc}. 

\section{Simple example}
\label{sec:ex}
Before moving on to the formulation applied to any MPGs for TIs and TSCs, we give a simple example to draw what we want to do in this paper. 
In this section, $\s_{\mu=0,x,y,z}$ and $\tau_{\mu=0,x,y,z}$ represent 2 by 2 Pauli matrices. 

Let us consider $1d$ time-reversal symmetry (TRS)-broken odd-parity superconductors (SCs), i.e., SCs in which the inversion transformation keeps the normal states invariant but flips the sign of the gap functions.
The topological nature of bulk is described by the Dirac Hamiltonian
\begin{align}
H(\hat k_x)=-i\p_x \g_1+m\G_0, \qquad 
\g_1^2=\G_0^2=1, \qquad \{\g_1,\G_0\}=0, 
\label{eq:1dDirac}
\end{align}
where $\hat k_x=-i\p_x$, and $m$ is a mass parameter.
The symbols $\g_1$ and $\G_0$ are gamma matrices with the anticommutation relation. 
The particle-hole and inversion symmetries are written as 
\begin{align}
&\hat CH(\hat k_x)\hat C^{-1}=-H(-\hat k_x), \qquad \hat C^2=1, \nonumber \\
&\hat I H(\hat k_x) \hat I^{-1}=H(-\hat k_x), \qquad \hat I^2=1, \nonumber \\
&\hat C \hat I=-\hat I\hat C. 
\label{eq:1d_sym_alg}
\end{align}
Here, $\hat I$ is the unitary inversion operator, and $\hat C$ is the antiunitary particle-hole operator. 
The anticommutation relation between $\hat C$ and $\hat I$ is from the assumption of the odd parity of the superconducting gap function. 
(See Sec.~\ref{sec:SC} for the detail.)
Note that the following algebraic relations hold 
\begin{align}
&\hat C\g_1\hat C^{-1}=\g_1, \qquad \hat I\g_1\hat I^{-1}=-\g_1, \nonumber \\
&\hat C\G_0\hat C^{-1}=-\G_0, \qquad \hat I\G_0\hat I^{-1}=\G_0. \label{eq:1dAlg_Gamma}
\end{align}
To carry out the topological classification of the mass term $m\G_0$, we introduce the modified operator $\tilde I= i \g_1 \hat I$ that behaves as an onsite chiral symmetry 
\begin{align}
\tilde I H(\hat k_x) \tilde I^{-1}=-H(\hat k_x), \qquad \tilde I^2=1, 
\end{align}
with the algebra among symmetry operators 
\begin{align}
\tilde I^2=1, \qquad \hat C \tilde I = \tilde I \hat C. 
\end{align}
The operators $\hat C$ and $\tilde I$ compose the class BDI in the Altland-Zirnbauer (AZ) symmetry classes~\cite{AZ}.
Thus, the mass terms $m\G_0$ are classified by integers, which we denote the $K$-group by $K=\Z$~\cite{RyuTenFold, KitaevPeriodic}.
The generator model is given by 
\begin{align}
H(\hat k_x)=-i\p_x\tau_y+m\tau_z, \qquad 
\hat C=\tau_x{\cal K}, \qquad 
\tilde I=\tau_x.
\label{eq:1dSC}
\end{align}
Here, ${\cal K}$ is the complex conjugation. 
Given a Dirac Hamiltonian with the modified inversion symmetry $\tilde I$, the $\Z$ index is given by the $1d$ winding number
\begin{align}
w_{1d}[H(\hat k_x)]=\frac{1}{2i} \tr[\g_1 \G_0 \tilde I], 
\label{eq:1dwinding}
\end{align}
which we will use later.~\footnote{
\label{fn:number_zeros}
The 1d winding number $w_{1d}$ of the Hamiltonian (\ref{eq:1dDirac}) for the chiral symmetry operator $\tilde I$ is given by 
\begin{align*}
    w_{1d} = \frac{1}{4 \pi i} \int_{-\infty}^{\infty} dk_x \tr[\tilde I H(k_x)^{-1} \partial_{k_x} H(k_x)]
    = \frac{-{\rm sgn}(m)}{4\pi i}\tr[\g_1 \G_0 \tilde I].
\end{align*}
The formula (\ref{eq:1dwinding}) gives the difference of the winding numbers when the mass parameter $m$ crosses the zero from the positive to the negative. 
}

Not every integer of the $K$-group $K$ represents a 1st-order TSC, the Kitaev chain. 
As is shown shortly, even integers of $K$ are 2nd-order TSCs, i.e., they are equivalent to atomic insulators localized at the inversion center. 
To see this, we add a symmetry-allowed Jackiw-Rebbi kink term $M(x)\G_1$ to \eqref{eq:1dDirac} as in 
\begin{align}
&H_{\rm kink}(\hat k_x,x)=
-i\p_x\g_1+M(x)\G_1+m\G_0, \nonumber \\
&\g_1^2=\G_0^2=\G_1^2=1, \nonumber \\
&\{\g_1,\G_0\}=\{\g_1,\G_1\}=\{\G_0,\G_1\}=0. 
\label{eq:1DDiracKink}
\end{align}
We also assume the anticommutation relation $\{\Gamma_0, \tilde I\}=0$ to be compatible with the sign change of the mass term $M(x)$ at $x=0$. 
The particle-hole and inversion symmetries are written as 
\begin{align}
&\hat C H_{\rm kink}(\hat k_x,x)\hat C^{-1}=-H_{\rm kink}(-\hat k_x,x), \qquad \hat C^2=1, \nonumber \\
&\hat I H_{\rm kink}(\hat k_x,x)\hat I^{-1}=H_{\rm kink}(-\hat k_x,-x), \qquad \hat I^2=1, \nonumber \\
&\hat C\hat I=-\hat I\hat C. 
\label{eq:1d_kink_sym_alg}
\end{align}
Here, $M(-x)=-M(x)$ is imposed from the inversion symmetry.~\footnote{As long as we are concerned with the bound states localized at the inversion center, the configuration of $M(x)$ is assumed as $M(x) = x$ (or $M(x) = -x$). 
Therefore, the kink term $M(x) \G_1$ behaves like an additional kinetic term of the Dirac Hamiltonian.
}
The key observation is the one-to-one correspondence between the Jackiw-Rebbi low-energy bound states of $H_{\rm kink}(\hat k_x,x)$ and atomic insulators at the inversion center.
The Jackiw-Rebbi Hamiltonian $-i \partial_x \gamma_1 + M(x) \Gamma_1$ hosts zero modes localized at the inversion center $x=0$. 
The uniform mass term $m \Gamma_0$ induces the finite energy to those zero modes, resulting in the localized modes at $x=0$ with finite positive or negative energies, which resemble atomic states. 
(See footnote 3 for the quantitative detail. ~\footnote{
The existence of the localized modes can be understood as follows. 
For simplicity, we set $M(x \to \infty)>0$. 
For the wave function of the form $\phi(x) = \eta e^{-\int^x M(x') dx'}$, where $\eta$ represents the internal degrees of freedom, the Jackiw-Rebbi Hamiltonian $-i \partial_x \gamma_1 + M(x) \Gamma_1$ hosts the zero modes determined by $(i \gamma_1 + \Gamma_1)\eta=0$. 
Due to the chiral symmetry $\{-i \partial_x \gamma_1, + M(x) \Gamma_1, \Gamma_0\}$ of the Jackiw-Rebbi Hamiltonian, the zero modes can be chosen as simultaneous eigenstates of $\Gamma_0$, $\Gamma_0 \eta = \pm \eta$, called chirality.  
The chiral index $N$ is defined as the difference of numbers of chiral zero modes with positive and negative chiralities, which is given by 
\begin{align*}
    N = \tr \left[ 
    \left(\frac{1+i \gamma_1 \Gamma_1}{2}\right) \Gamma_0
    \right] = 
    -\frac{1}{2i}\tr [\gamma_1 \Gamma_1 \Gamma_0]
    \in \Z, 
\end{align*}
where $(1+i \gamma_1 \Gamma_1)/2$ is the projector onto the subspace of zero modes. 
We note that only the $N$ chiral zero modes are stable under the chiral symmetry by $\Gamma_0$. 
The uniform mass $m \Gamma_0$ in (\ref{eq:1DDiracKink}) gives the chiral zero modes the finite energy $\pm m$ with the sign determined by the chirality $\Gamma_0 = \pm 1$. 
})

The aforementioned one-to-one correspondence is systematically constructed by the dimensional isomorphic mapping in the $K$-theory~\cite{Teo-Kane}. 
We denote the classification of atomic insulators at the inversion center by $K_{\rm AI}$. 
The group $K_{\rm AI}$ is represented by a $0d$ Hamiltonian $H_{0d}$ with the particle-hole and inversion symmetries, 
\begin{align}
&\hat C_{0d} H_{0d} \hat C_{0d}^{-1}=-H_{0d}, \qquad \hat C_{0d}^2=1, \nonumber\\
&\hat I_{0d} H_{0d} \hat I_{0d}^{-1}=H_{0d}, \qquad \hat I_{0d}^2=1, \nonumber \\
&\hat C_{0d} \hat I_{0d} = -\hat I_{0d} \hat C_{0d}. 
\end{align}
From $\hat C_{0d}\hat I_{0d}=-\hat I_{0d}\hat C_{0d}$, the particle-hole symmetry (PHS) operator $\hat C_{0d}$ exchanges the positive and negative eigenvalues of $\hat I_{0d}$, meaning that the AZ class for the Hamiltonian $H_{0d}$ is effectively the class A. 
Thus, we find that the $K$-group $K_{\rm AI}$ is $K_{\rm AI} = \Z$ and is generated by the Hamiltonian $H_{0d}=m \s_z$ with symmetry operators $\hat C_{0d}=\s_x {\cal K}$ and $\hat I_{0d} = \s_z$.
Given a $0d$ Hamiltonian $H_{0d}$ with symmetry operators $\hat C_{0d}$ and $\hat I_{0d}$, the dimensional raising isomorphic map to the kink Hamiltonian is given by
\begin{align}
H_{\rm kink}(\hat k_x,x)
:=
-i\p_x\tau_y+M(x)\tau_x+H_{0d} \tau_z, \qquad 
\tilde C=\tilde C_{0d}\tau_z, \qquad 
\hat I = \hat I_{0d}\tau_z.
\label{eq:1DDiracKink_2}
\end{align}
Notice that the Hamiltonian $H_{\rm kink}$ and the symmetry operators $\hat C$ and $\hat I$ obtained by this way automatically satisfy the desired symmetry algebra (\ref{eq:1d_kink_sym_alg}). 
In particular, the generator model of $K_{\rm AI}=\Z$ is  mapped as 
\begin{align}
H_{\rm kink}(\hat k_x,x)
= 
-i\p_x\tau_y+M(x)\tau_x+m\s_z\tau_z, \qquad 
\hat C=\s_x\tau_z{\cal K}, \qquad 
\hat I = \s_z\tau_z.
\label{eq:1d_KAI_GeneratorModel}
\end{align}

The final step is to determine which element of the $K$-group $K$ admits a kink term $M(x)\G_1$, which is given by the homomorphism $f: K_{\rm AI} \to K$ defined through the identification of $H_{0d}$ and $H_{\rm kink}(\hat k_x,x)$.
By setting the kink mass $M(x)$ to be zero, the Dirac Hamiltonian (\ref{eq:1DDiracKink_2}) is regarded as one without $x$-dependence, which defines the homomorphism $f$. 
For the generator model \eqref{eq:1d_KAI_GeneratorModel}, the modified inversion operator is $\tilde I = i \g_1 \hat I = -\tau_x \s_z$, and the $1d$ winding number $w_{1d}$ defined in \eqref{eq:1dwinding} is given by 
\begin{align}
w_{1d}\left[ \tilde H_{\rm kink}(\hat k_x,x)|_{M(x) \to 0} \right] 
= 
\frac{1}{2i} \tr[\tau_y \times \s_z \tau_z \times (-\tau_x \s_z)]=-2.
\end{align}
This means that within the $K$-group $K=\Z$, Hamiltonians admitting the kink term belong to even integers $\im f = 2\Z \subset \Z$, resulting in 2nd-order TSCs (atomic insulators). 
Let us write the abelian group of the 2nd-order TSCs by $K':=\im f$. 
We have the subgroup structure $K' \subset K$ and conclude that the classification of 1st-order TSCs is given by the quotient $K/K' = \Z/2\Z$.

\section{Formulation}
\label{sec:formulation}
In this section, generalizing the strategy illustrated in the previous section, we formulate how to compute the abelian group $K$ of all TIs/TSCs and the abelian group $K'''$ composed only of 4th-order TIs/TSCs in three space dimensions. 
The generalization to any space dimensions is straightforward.

\subsection{The entire $K$-group $K$ for TIs/TSCs}
\label{sec:K}
Let $H$ be a $3d$ Dirac Hamiltonian with a uniform mass,  
\begin{align}
&H(\bk)
=k_x\g_1+k_2\g_2+k_3\g_3+m\G_0, \nonumber \\
&\{\g_i,\g_j\}=2\delta_{ij}, \qquad \G_0^2=1, \qquad \{\g_i,\G_0\}=0.
\end{align}
Let $G$ be an MPG equipped with the data $(O_g, \phi_g,c_g,z_{g,h})$ for $g,h \in G$. 
The matrix $O_g \in O(3)$ represents how the group $G$ acts on the real-space coordinate as $\bm{x} \mapsto O_g \bm{x}$ for $g \in G$.
The homomorphisms $\phi, c: G \to \Z_2=\{\pm 1\}$ indicate unitary/antiunitary and symmetry/antisymmetry of the group element $g \in G$, respectively.
The symmetry constraint relations are written as 
\begin{align}
\hat g H(\bk) \hat g^{-1}
= 
c_g H(\phi_g O_g\bk), \qquad 
\hat g i \hat g^{-1} = \phi_g i, \qquad 
g \in G.
\label{eq:bulk_g_sym}
\end{align}
The set of $U(1)$ phases $z_{g,h} \in U(1)$ for $g,h \in G,$ specifies the factor system of a projective representation
\begin{align}
\hat g \hat h = z_{g,h} \wh{gh}, \qquad g,h \in G.
\end{align}
For gamma matrices $\g_i$s and $\G_0$, the symmetry (\ref{eq:bulk_g_sym}) is written as 
\begin{align}
\hat g \bm{\g} \hat g^{-1} =\phi_g c_g O_g^{-1} \bm{\g}, \qquad 
\hat g \G_0 \hat g^{-1}=c_g \G_0. 
\label{eq:g-sym}
\end{align}

For an $SO(3)$ rotation $R = (\bm{n},\theta)$ with the counterclockwise rotation about $\bm{n}$-axis by the angle $\theta$, we have the key equality 
\begin{align}
O_R \bm{\g} =q_R\bm{\g}q_R^{-1}, 
\end{align}
with $q_R$ the unitary operator canonically defined by 
\begin{align}
q_R=e^{\frac{\theta}{2} (n_1\g_2\g_3+n_2\g_3\g_1+n_3\g_1\g_2)} 
\end{align}
irrespective to representations of the gamma matrices, where $\bm{n} = (n_1,n_2,n_3)$ is a unit vector.~\footnote{
In generic $d$-space dimensions, for an $SO(d)$ rotation $R=\exp\left( \frac{i}{2} \theta_{ij} L_{ij} \right)$, in which $[L_{ij}]_{kl}=-i(\delta_{ik}\delta_{jl}-\delta_{jk}\delta_{il})$ are the generators of $SO(d)$, a $Spin(d)$ rotation is given by $q_R=\exp\left( \frac{i}{2} \theta_{ij} \Sigma_{ij} \right)$, 
where $\Sigma_{ij}=\frac{-i}{4}[\g_i,\g_j]$ are the generators of $Spin(d)$ rotations. 
}
The operator $q_R$ can be used to make $\hat g$ onsite~\cite{CornfeldPointGroup}.
We further introduce a homomorphism $s:G \to \Z_2=\{\pm 1\}$ by $s_g:=\det[O_g] \in \{\pm 1\}$ specifying if $g\in G$ preserves the orientation or not. 
Let $R_g$ be the $SO(3)$ part of $O_g \in O(3)$, i.e., $R_g = O_g$ for $s_g=1$, and $R_g = I O_g$ for $s_g=-1$, where $I={\rm diag}(-1,-1,-1)$ is the space inversion.
We introduce the new symmetry operator $\tilde g$ defined by 
\begin{align}
\tilde g:=
(\g_1\g_2\g_3)^{(1-s_g)/2} q_{R_g} \hat g.
\end{align}
The new operators now represent onsite symmetry
\begin{align}
\tilde g H(\bk) \tilde g^{-1}=c_g s_g H(\phi_g \bk),\qquad 
\tilde g i \tilde g^{-1} = \phi_g i, \qquad 
g \in G.
\end{align}
Equivalently, for gamma matrices, 
\begin{align}
\tilde g \bm{\g} \tilde g^{-1} = c_g s_g\phi_g\bm{\g}, \qquad 
\tilde g \G_0 \tilde g^{-1}=c_g s_g \G_0,\quad g \in G.
\end{align}
It is to be noted that orientation-reversing symmetry operators (i.e. for operators with $\phi_g=1$, $s_g=-1$, and $c_g=1$) behave as chiral symmetry. 
Also, we note that the new operators $\tilde g$ obey a different factor system from that for $\hat g$s: 
From a straightforward calculation, the factor system $\tilde z_{g,h}$ of $\tilde g$s defined by $\tilde g \tilde h = \tilde z_{g,h} \wt{gh}$ is~\footnote{
Use the relations $q_{R_g} (\g_1\g_2\g_3) = (\g_1\g_2\g_3) q_{R_g}$ and $\tilde g q_{R_h}=q_{R_h}\tilde g$.}
\begin{align}
&\tilde z_{g,h}=z'_{g,h} \times (-1)^{\frac{1-c_g\phi_g}{2}\frac{1-s_h}{2}}\times z_{g,h} , 
\label{eq:factor_system_ztilde}
\\
&z'_{g,h}:= q_{R_h} q_{R_g} q_{R_{gh}}^{-1} \in \{\pm 1\}.
\label{eq:change_factor_system}
\end{align}
Here we have introduced the factor system $z'_{g,h} (=(z'_{g,h})^{-1})$ of the $Spin(3)$ rotation matrices $q_R$ with the right group action, which is determined for a fixed choice of $SO(3)$-rotation parameters $(\bm{n}_g,\theta_g)$ for $g \in G$. 
We will use $z'_{g,h}$ frequently later.

Since the MPG $G$ is now represented as onsite symmetry, one can apply the Wigner criteria to symmetry operators $\{\tilde g\}_{g \in G}$ to get the $K$-group $K$ classifying the mass term $m\G_0$~\cite{KS_Atiyah}. 
We decompose the group $G$ into subsets as 
\begin{align}
G=G_0 \coprod aG_0 \coprod bG_0 \coprod ab G_0, 
\end{align}
where 
\begin{align}
&G_0 = \{g \in G|\phi_g=1, c_g s_g =1 \}, \\
&{}^\exists a \in G, \quad -\phi_a=c_a s_a=1,\\
&{}^\exists b \in G, \quad -\phi_b=-c_b s_b=1 ,\\
&{}^\exists ab \in G, \quad \phi_{ab}=-c_{ab} s_{ab}=1.
\end{align}
(Here, $ab$ is not necessarily the group product of $a$ and $b$, but represents a single element of $G$.)
For an irrep $\alpha$ of $G_0$ with the factor system $\tilde z_{g,h}$, the Winger criteria for the operators $\tilde a$ and $\tilde b$ are defined by 
\begin{align}
&W_{\alpha}^T
=
\frac{1}{|G_0|} \sum_{g \in aG_0} \tilde z_{g,g}\tilde \chi_{\alpha}(g^2) \in \{0,\pm 1\}, \label{eq:wigner_t} \\
&W_{\alpha}^C
=
\frac{1}{|G_0|} \sum_{g \in bG_0} \tilde z_{g,g}\tilde \chi_{\alpha}(g^2) \in \{0,\pm 1\}, \label{eq:wigner_c}
\end{align}
where $\tilde \chi_\alpha(g)$ is the irreducible character of $\alpha$. 
We also introduce the orthogonal test for the operator $\wt{ab}$ by  
\begin{align}
&O_{\alpha\alpha}^{\G}
= 
\frac{1}{|G_0|} \sum_{g \in G_0} \left[ \frac{\tilde z_{g,ab}}{\tilde z_{ab,(ab)^{-1}g ab}} \tilde \chi_{\alpha}((ab)^{-1}g ab) \right]^* \tilde \chi_{\alpha}(g) \in \{0,1\}.
\label{eq:ortho}
\end{align}
There are 19 patterns of possible combinations of $W^T_\alpha,W^C_\alpha,O^\G_{\alpha\alpha}$, which are shown in Table~\ref{tab:wigner_az}, and we name them 
A, AI, AII, A$_T$, D, C, A$_C$, AIII, A$_\G$, A$_{T,C}$, AIII$_T$, AI$_C$, BDI, D$_T$, DIII, AII$_C$, CII, C$_T$, and CI.
We call them effective AZ (EAZ) class of $\alpha$. 
Given an EAZ class, the topological classification of the mass term $m\G_0$ can be read off from the periodic table of TIs/TSCs~\cite{RyuTenFold, KitaevPeriodic} shown in Table~\ref{tab:periodictable}. 
Summing up all the contributions from irreps of $G_0$, the $K$-group $K$ of TIs/TSCs is determined. 

\begin{table}
	\centering
	\caption{The relationship among the Wigner criteria $W^T_{\alpha}, W^C_{\alpha}$, orthogonal test $O^{\Gamma}_{\alpha\alpha}$, EAZ classes, and band structures. }
	\label{tab:wigner_az}
	\footnotesize
\begin{align*}
\begin{array}{cccccc}
& {\rm EAZ} & {\rm Band\ str.}\\
\hline \hline
& {\rm A} & 
\begin{minipage}{20mm}
        \scalebox{0.5}{\includegraphics[width=40mm,clip]{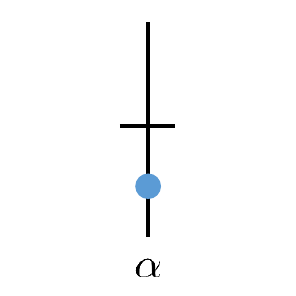}}
    \end{minipage}\\
\hline \hline
\medskip \\
W^T_{\alpha} & {\rm EAZ} & {\rm Band\ str.}\\
\hline \hline
1 & {\rm AI} & 
\begin{minipage}{20mm}
        \scalebox{0.5}{\includegraphics[width=40mm,clip]{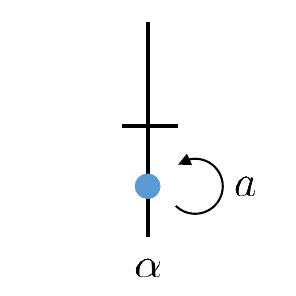}}
    \end{minipage}\\
\hline
-1 & {\rm AII} & 
\begin{minipage}{20mm}
        \scalebox{0.5}{\includegraphics[width=40mm,clip]{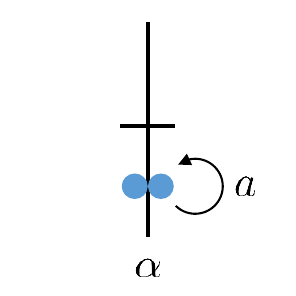}}
    \end{minipage}\\
    \hline
0 & {\rm A}_T & 
\begin{minipage}{20mm}
        \scalebox{0.5}{\includegraphics[width=40mm,clip]{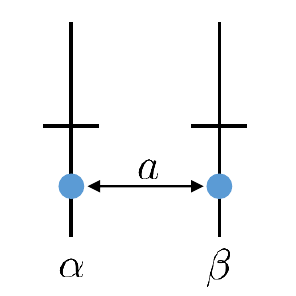}}
    \end{minipage}\\
\hline \hline
\medskip \\
W^C_{\alpha} & {\rm EAZ} & {\rm Band\ str.}\\
\hline \hline
1 & {\rm D} & 
\begin{minipage}{20mm}
        \scalebox{0.5}{\includegraphics[width=40mm,clip]{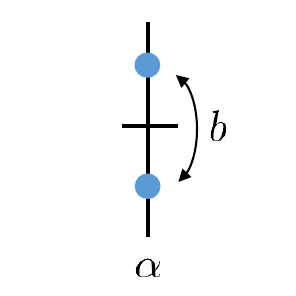}}
    \end{minipage}\\
\hline
-1 & {\rm C} & 
\begin{minipage}{20mm}
        \scalebox{0.5}{\includegraphics[width=40mm,clip]{d}}
    \end{minipage}\\
    \hline
0 & {\rm A}_C & 
\begin{minipage}{20mm}
        \scalebox{0.5}{\includegraphics[width=40mm,clip]{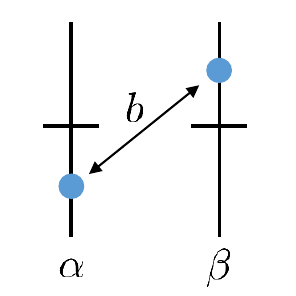}}
    \end{minipage}\\
\hline \hline
\medskip \\
O^{\Gamma}_{\alpha\alpha} & {\rm EAZ} & {\rm Band\ str.}\\
\hline \hline
1 & {\rm AIII} & 
\begin{minipage}{20mm}
        \scalebox{0.5}{\includegraphics[width=40mm,clip]{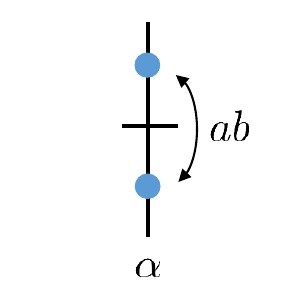}}
    \end{minipage}\\
\hline
0 & {\rm A}_\Gamma & 
\begin{minipage}{20mm}
        \scalebox{0.5}{\includegraphics[width=40mm,clip]{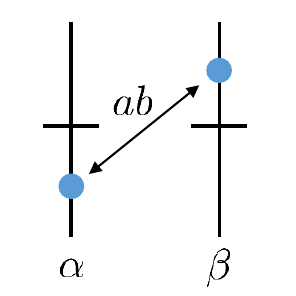}}
    \end{minipage}\\
\hline \hline
\medskip \\
\medskip \\
\ \\
\end{array} && && 
\begin{array}{ccccccccc}
W^T_{\alpha} & W^C_{\alpha} & O^{\Gamma}_{\alpha\alpha} & {\rm EAZ} & {\rm Band\ str.}\\
\hline \hline
0&0&0&{\rm A}_{T,C}& 
\begin{minipage}{25mm}
        \scalebox{0.5}{\includegraphics[width=55mm,clip]{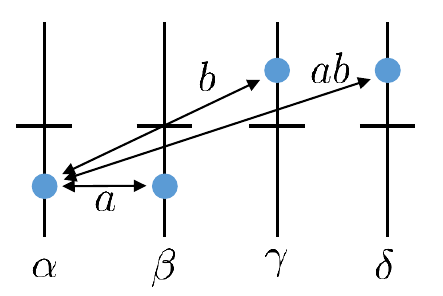}}
    \end{minipage}\\ 
    \hline
0&0&1&{\rm AIII}_T & 
\begin{minipage}{20mm}
        \scalebox{0.5}{\includegraphics[width=40mm,clip]{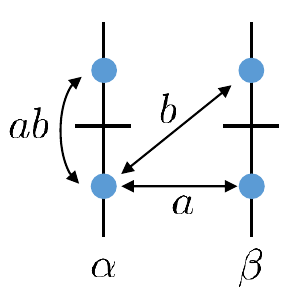}}
    \end{minipage}\\
    \hline
1&0&0&{\rm AI}_C & 
\begin{minipage}{20mm}
        \scalebox{0.5}{\includegraphics[width=40mm,clip]{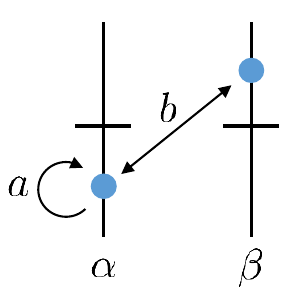}}
    \end{minipage}\\
    \hline
1&1&1&{\rm BDI}& 
\begin{minipage}{20mm}
        \scalebox{0.5}{\includegraphics[width=40mm,clip]{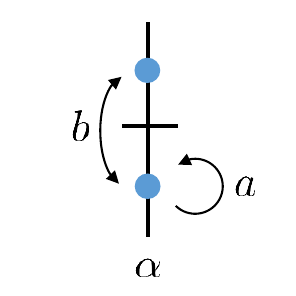}}
    \end{minipage}\\
    \hline
0&1&0&{\rm D}_T & 
\begin{minipage}{20mm}
        \scalebox{0.5}{\includegraphics[width=40mm,clip]{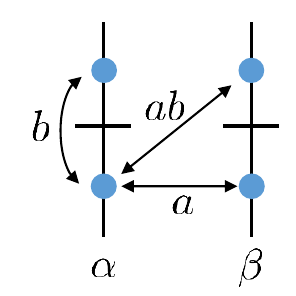}}
    \end{minipage}\\
    \hline
-1&1&1&{\rm DIII} & 
\begin{minipage}{20mm}
        \scalebox{0.5}{\includegraphics[width=40mm,clip]{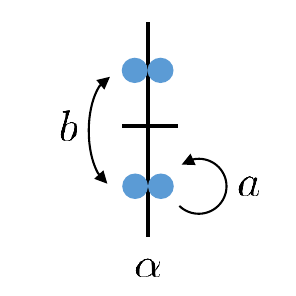}}
    \end{minipage}\\
    \hline
-1&0&0&{\rm AII}_C & 
\begin{minipage}{20mm}
        \scalebox{0.5}{\includegraphics[width=40mm,clip]{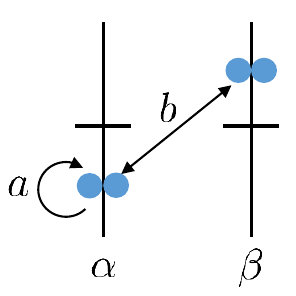}}
    \end{minipage}\\
    \hline
-1&-1&1&{\rm CII}&
\begin{minipage}{20mm}
        \scalebox{0.5}{\includegraphics[width=40mm,clip]{d3_both}}
    \end{minipage}\\
    \hline
0&-1&0&{\rm C}_T & 
\begin{minipage}{20mm}
        \scalebox{0.5}{\includegraphics[width=40mm,clip]{d_both}}
    \end{minipage}\\
    \hline
1&-1&1&{\rm CI} & 
\begin{minipage}{20mm}
        \scalebox{0.5}{\includegraphics[width=40mm,clip]{bd1_both}}
    \end{minipage}\\
\hline \hline
\vspace{80pt}\\
\end{array}
\end{align*}
\end{table}

\begin{table}
\caption{The periodic table of TIs/TSCs.}
\label{tab:periodictable}
$$
\begin{array}{l|cccccccc}
\mbox{EAZ class} & d=0 & d=1 & d=2 & d=3 & d=4 & d=5 & d=6 & d=7\\
\hline
{\rm A},{\rm A}_T,{\rm A}_C,{\rm A}_\Gamma,{\rm A}_{T,C}&\Z&0&\Z&0&\Z&0&\Z&0\\
{\rm AIII},{\rm AIII}_T&0&\Z&0&\Z&0&\Z&0&\Z\\
\hline
{\rm AI},{\rm AI}_C&\Z&0&0&0&2\Z&0&\Z_2&\Z_2\\
{\rm BDI}&\Z_2&\Z&0&0&0&2\Z&0&\Z_2\\
{\rm D},{\rm D}_T&\Z_2&\Z_2&\Z&0&0&0&2\Z&0\\
{\rm DIII}&0&\Z_2&\Z_2&\Z&0&0&0&2\Z\\
{\rm AII},{\rm AII}_C&2\Z&0&\Z_2&\Z_2&\Z&0&0&0\\
{\rm CII}&0&2\Z&0&\Z_2&\Z_2&\Z&0&0\\
{\rm C},{\rm C}_T&0&0&2\Z&0&\Z_2&\Z_2&\Z&0\\
{\rm CI}&0&0&0&2\Z&0&\Z_2&\Z_2&\Z\\
\end{array}
$$
\end{table}

\subsection{Higher-order TIs/TSCs}
Not every element of the $K$-group $K$ hosts a gapless surface state since some elements of $K$ may be compatible with spatially varying masses, which we call the defect mass terms, that induce a finite energy gap to the surface state. 
The relationship among the $K$-group $K$ and surface states is best described by the structure of subgroups~\cite{HermeleBuilding, LukaHigherOrder} 
\begin{align}
0 \subset K''' \subset K'' \subset K' \subset K 
\label{eq:subgroup_str}
\end{align}
for three space dimensions, where $K^{(n)}$ denotes the abelian group composed of Dirac Hamiltonians that admit at least $n$ additional defect mass terms~\cite{LukaHigherOrder, KhalafPRX18AII} as in 
\begin{align}
&H(\hat \bk,\bm{r})= -i \bm{\g} \cdot \bm{\p} + m \G_0 + \sum_{j=1}^{n} M_j(\bm{x}) \G_j, \\
&\{ \g_i, \g_j\} = 2\delta_{ij}, \qquad 
\{\G_i,\G_j\} = 2\delta_{ij}, \qquad 
\{\g_i,\G_j\}=0.
\end{align}
Here, we have introduced $\hat \bk = -i \bm{\p}$. 
The additional defect mass terms $M_j(\bm{x})\G_j$ decrease the dimensionality of the surface Dirac fermion. 
By design, the quotient group $K^{(n-1)}/K^{(n)}$ represents Dirac Hamiltonians that admit at most $n$ defect masses, which means the surface Dirac fermion is constraint on a $(3-n)$-dimensional subregion, and such phases are called $n$th-order TIs/TSCs. 
In particular, the group $K'''$ of 4th-order TIs/TSCs, which are composed of Dirac Hamiltonians with $3$ defect masses, represent no surface states but a bound state localized at the point group center (i.e., $\bm{r}=\bm{0}$).~\footnote{ 
Any point group has at least one fixed point under the action of the corresponding symmetry operation, and we call the fixed point the point group center.}

See Ref.~\cite{LukaHigherOrder} for explicit subgroups (\ref{eq:subgroup_str}) for an additional order-two MPG symmetry for the ten-fold AZ symmetry classes.
For generic MPG symmetry, it has not yet known how to compute all the subgroups in (\ref{eq:subgroup_str}). 
Nevertheless, in addition to the entire $K$-group $K$, one can compute the group $K'''$ of 4th-order TIs/TSCs in a canonical way, which is developed in the rest of this section. 
Therefore, we have the quotient group $K/K'''$, the topological classification of surface states.

\subsection{Dirac Hamiltonian with a hedgehog mass}
Let $K_{\rm AI}$ be the abelian group generated by atomic insulators exactly at the point group center. 
The group $K_{\rm AI}$ is nothing but the abelian group generated by 0-dimensional Hamiltonians with the symmetry group $G$ with the homomorphisms $\phi_g, c_g$ and the factor system $z_{g,h}$. 
To compute the group $K'''$, we first construct the homomorphism from the group $K_{\rm AI}$ to the group $K$ of TIs/TSCs as follows. 
There is an isomorphism between the group $K_{\rm AI}$ and the group of $3d$ Dirac Hamiltonians with a hedgehog-mass potential,  
\begin{align}
H(\hat \bk,\bm{r}) = \sum_{\mu=1}^3 -i \p_\mu \g_\mu + m \Gamma_0 + \sum_{\mu=1}^3 M_\mu(\bm{r}) \Gamma_\mu, 
\label{eq:3dDirac_hedgehog}
\end{align}
with a unit winding number 
\begin{align}
\frac{1}{8\pi} \int_{|\bm{r}| \to \infty} dS \epsilon_{ij} \epsilon_{\mu\nu\rho} \hat M_\mu(\bm{r}) \p_i \hat M_\nu(\bm{r}) \p_j \hat M_\rho(\bm{r}) = 1, 
\end{align}
where $\hat M_\mu(\bm{r})= M_\mu(\bm{r})/|\bm{M}(\bm{r})|$. 
Here, the amplitude $|\bm{M}(\bm{r})|$ of the mass vector $\bm{M}(\bm{r})=\big(M_1(\bm{r}),M_2(\bm{r}),M_3(\bm{r})\big)$ is supposed to be finite at the infinity $|\bm{r}| \to \infty$ so that $H(\hat \bk,\bm{r})$ hosts an exponentially localized bound state around the point group center. 
This relationship is, in fact, an isomorphism: 
As we will see shortly, given an atomic insulator $H_{0d}$ with symmetry constrained by $G$, one can construct the $3d$ Dirac Hamiltonian with a hedgehog mass in the form (\ref{eq:3dDirac_hedgehog}). 
Conversely, a $3d$ Dirac Hamiltonian with a hedgehog mass with a unit winding number, we have a $0d$ bound state that is symmetric under the group $G$. 

Neglecting mass vectors by setting $\bm{M}(\bm{r}) \equiv \bm{0}$ defines the homomorphism $f: K_{\rm AI} \to K$. 
It should be noted that not every atomic insulator of $K_{\rm AI}$ is pinned at the point group center since some sets of atomic orbitals of $K_{\rm AI}$ can go far away without breaking the $G$ symmetry, and such combinations of atomic orbitals should be zero in the target group $K$.~\footnote{ 
This is because nonzero elements of $K$ represent either a surface state (i.e., a 1st, 2nd, or 3rd-order TI/TSC) of the 3D ball or a bound state (i.e., 4th-order TI/TSC) localized and pinned at the point group center.}
Thus, the group $K'''$ representing bound states pinned at the point group center is given by the image of $f$, $K''':= \im [f: K_{\rm AI} \to K] \subset K$. 

As far as the topological classification is concerned, the hedgehog-mass vector with a unit winding number can be set as $M_\mu(\bm{r}) = r_\mu$. 
In doing so, the Hamiltonian becomes a Dirac Hamiltonian with six kinetic-like parameters $(\bk,\bm{r})$ as 
\begin{align}
H_{3d}(\bm{k},\bm{r})
= 
\bm{k} \cdot \bm{\g} + \bm{r} \cdot \bm{\Gamma} + m \Gamma_0. 
\end{align}
The symmetry constraint is written as 
\begin{align}
\hat g H_{3d}(\bk,\bm{r}) \hat g^{-1}
= 
c_g H_{3d}(\phi_g O_g\bk, O_g \bm{r}).  \qquad 
\hat g \hat h = z_{g,h} \wh{gh}.  
\label{eq:3d_Hamiltonian_mass_hedgehog_sym}
\end{align}
Equivalently, for gamma matrices, 
\begin{align}
\hat g \bm{\g} \hat g^{-1} =\phi_g c_g O_g^{-1} \bm{\g}, \qquad 
\hat g \bm{\Gamma} \hat g^{-1} = c_g O_g^{-1} \bm{\Gamma}, \qquad 
\hat g \Gamma_0 \hat g^{-1}=c_g \Gamma_0.
\end{align}
As we did in Sec.~\ref{sec:K}, we introduce new point group operators $\check g$ as 
\begin{align}
&\check g := (i \g_1\g_2\g_3\G_1\G_2\G_3)^{\frac{1-s_g}{2}} \times q_{R_g} \times  Q_{R_g} \times \hat g, 
\label{eq:sym_op_ghat_to_gbar}
\\
&q_R=
e^{\frac{\theta}{2} (n_1\g_2\g_3+n_2\g_3\g_1+n_3\g_1\g_2)}, \\
&Q_R=
e^{\frac{\theta}{2} (n_1\G_2\G_3+n_2\G_3\G_1+n_3\G_1\G_2)}, 
\end{align}
so that $\check g$ acts on the Hamiltonian as onsite symmetry, 
\begin{align}
\check g H_{3d}(\bk,\bm{r}) \check g^{-1}
= 
c_g H_{3d}(\phi_g \bk, \bm{r}).
\end{align}
From a straightforward calculation, we find that $\check g \check h = z_{g,h} \widecheck{gh}$, that is, the factor system of $\check g$s is the same as $z_{g,h}$ for $\hat g$s, as expected. 

Since there exist the same number of ``$k$-type'' and ``$r$-type'' coordinates, the topological classification of the mass term $m \G_0$ is recast as that for $0d$ Hamiltonians $H_{0d}$ with the same symmetry class, which is dubbed ``a defect gapless state as a boundary state''~\cite{Teo-Kane}. 
The explicit construction of the dimensional raising isomorphism is as follows. 
Let 
\begin{align}
G = G_0^{\rm AI} \coprod \underline{a} G_0^{\rm AI} \coprod \underline{b} G_0^{\rm AI} \coprod \underline{ab} G_0^{\rm AI}
\end{align}
be the decomposition of the group $G$ associated with the homomorphisms $\phi_g,c_g$.
Namely, 
\begin{align}
&G_0^{\rm AI}=\{g \in G|\phi_g=c_g=1\}, \\
&{}^\exists \underline{a}\in G,\quad -\phi_{\underline{a}}=c_{\underline{a}}=1, \\
&{}^\exists \underline{b}\in G,\quad -\phi_{\underline{b}}=-c_{\underline{b}}=1, \\
&{}^\exists \underline{ab}\in G,\quad -\phi_{\underline{ab}}=-c_{\underline{ab}}=1. 
\end{align}
Let $\beta$ be an irrep of $G_0^{\rm AI}$ with the factor system $z_{g,h}$.
The Wigner criteria and the orthogonal test for the irrep $\beta$ are defined as 
\begin{align}
&\underline{W}_{\beta}^T
=
\frac{1}{|G_0^{\rm AI}|} \sum_{g \in \underline{a}G_0^{\rm AI}} z_{g,g} \chi_{\beta}(g^2) \in \{0,\pm 1\}, 
\label{eq:winger_t_ai}
\\
&\underline{W}_{\beta}^C
=
\frac{1}{|G_0^{\rm AI}|} \sum_{g \in \underline{b} G_0^{\rm AI}} z_{g,g} \chi_{\beta}(g^2) \in \{0,\pm 1\}, 
\label{eq:winger_c_ai}
\\
&\underline{O}_{\beta \beta}^{\G}
= 
\frac{1}{|G_0^{\rm AI}|} \sum_{g \in G_0^{\rm AI}} \left[ \frac{z_{g,\underline{ab}}}{z_{\underline{ab},(\underline{ab})^{-1}g \underline{ab}}} \chi_{\beta}((\underline{ab})^{-1}g \underline{ab}) \right]^* \chi_{\beta}(g) \in \{0,1\}.
\label{eq:ortho_ai}
\end{align}
Here, $\chi_\beta(g)$ is the irreducible character of $\beta$. 
The triple $(\underline{W}^T_\beta,\underline{W}^C_\beta,\underline{O}^\G_{\beta\beta})$ determines the EAZ class of $\beta$. 
Let $D_\beta(g)$ for $g \in G_0^{\rm AI}$ be the representation matrices of the irrep $\beta$. 
We introduce the mapped representation of $\beta$ by $h \in G$ as 
\begin{align}
D_{h(\beta)}(g \in G_0^{\rm AI}):= 
\frac{z_{g,h}}{z_{h,h^{-1}gh}} \times 
\left\{\begin{array}{ll}
D_{\beta}(h^{-1}gh) & (\phi_h=1), \\
D_{\beta}(h^{-1}gh)^* & (\phi_h=-1). \\
\end{array}\right.
\end{align}
With this, the symmetry operators $\check{g}^{(0d)}$ for the $0d$ Hamiltonian $H_{0d}$ are given as, for elements with $\phi_g=c_g=1$, 
\begin{align}
\check{g}^{(0d)}|_{\phi_g=c_g=1} = 
\left\{\begin{array}{ll}
D_\beta(g) & \mbox{for A, AI}, \\
D_\beta(g)\oplus D_{\underline{a}(\beta)}(g) & \mbox{for AII, A$_T$}, \\
\begin{pmatrix}
D_\beta(g)\\
&D_{\underline{b}(\beta)}(g)
\end{pmatrix}_\tau& \mbox{for D, C, A$_C$, AI$_C$, BDI, CI}, \\
\begin{pmatrix}
D_\beta(g)\\
&D_{\underline{ab}(\beta)}(g)
\end{pmatrix}_\tau& \mbox{for AIII, A$_\G$}, \\
\begin{pmatrix}
D_\beta(g)\oplus D_{\underline{a}(\beta)}(g) \\
&D_{\underline{b}(\beta)}(g)\oplus D_{\underline{ab}(\beta)}(g)
\end{pmatrix}_\tau& \mbox{for A$_{T.C}$, AIII$_T$, D$_T$, DIII, AII$_C$, CII, C$_T$}.  \\
\end{array}\right.
\end{align}
For other combinations of $\phi_g$ and $c_g$, there are proper expressions of $\check g^{(0d)}$s, which we do not use later.  
Here, the subscript $\tau$ means the matrix of the particle-hole space. 
A representative $0d$ Hamiltonian $H_{0d}$ is given by $H_{0d}={\bf 1}$ when the group $G$ does not include an element $g$ with $c_g=-1$, or $H_{0d}={\bf 1} \otimes \tau_z$ when the group $G$ includes an element $g$ with $c_g=-1$, which we denote the two cases as 
\begin{align}
H_{0d} = 
    \left\{\begin{array}{ll}
{\bf 1} & (c_g =1, {}^\forall g \in G), \\
{\bf 1} \otimes \tau_z & (c_g = -1, {}^\exists g \in G). \\
\end{array}\right.
\end{align}

By design, $H_{0d}$ satisfies the symmetry constraints
\begin{align}
\check g^{(0d)} H_{0d} (\check g^{(0d)})^{-1} = c_g H_{0d}, \qquad 
\check g^{(0d)} \check h^{(0d)} = z_{g,h} \widecheck{gh}^{(0d)} 
\end{align}
for $g,h \in G$. 
Starting from the $0d$ Hamiltonian, we have the dimensional raising isomorphisms: 
\begin{align}
\left\{\begin{array}{ll}
&H_{1d}(k_1,r_1) := k_1 \s_y + r_1 \s_x + H_{0d} \s_z, \qquad 
\check{g}^{(1d)} = \check{g}^{(0d)} \s_z^{\frac{1-c_g}{2}}, \\
&\check{g}^{(1d)} H(k_1,r_1) (\check{g}^{(1d)})^{-1} = c_g H(\phi_g k_1,r_1), \qquad 
\check g^{(1d)} \check h^{(1d)} = z_{g,h} \widecheck{gh}^{(1d)}, 
\end{array}\right.
\end{align}
\begin{align}
\left\{\begin{array}{ll}
&H_{2d}(k_1,k_2,r_1,r_2) := k_2 s_y + r_2 s_x+ H_{1d}(k_1,r_1)  s_z, \qquad 
\check g^{(2d)} = \check g^{(1d)} s_z^{\frac{1-c_g}{2}}, \\
&\check{g}^{(2d)} H(k_1,k_2,r_1,r_2) (\check{g}^{(2d)})^{-1} = c_g H(\phi_g k_1,\phi_g k_2,r_1,r_2), \qquad 
\check g^{(2d)} \check h^{(2d)} = z_{g,h} \widecheck{gh}^{(2d)}, 
\end{array}\right.
\end{align}
\begin{align}
\left\{\begin{array}{ll}
&H_{3d}(\bk,\bm{r}) = k_3 \mu_y + r_3 \mu_x+ H_{2d}(k_1,k_2,r_1,r_2) \mu_z, \qquad 
\check g^{(3d)} = \check g^{(2d)} \mu_z^{\frac{1-c_g}{2}}, \\
&\check{g}^{(3d)} H(\bk,\bm{r}) (\check{g}^{(3d)})^{-1} = c_g H(\phi_g \bk,\bm{r}), \qquad 
\check g^{(3d)} \check h^{(3d)} = z_{g,h} \widecheck{gh}^{(3d)}. 
\end{array}\right.
\end{align}
Here, $\bm{\s}, \bm{s},\bm{\mu}$ are Pauli matrices, and we abbreviated $(k_1,k_2,k_3,r_1,r_2,r_3)$ by $(\bk,\bm{r})$. 
We eventually have the mapped $3d$ Hamiltonian $H_{3d}(\bk,\bm{r})$ with the hedgehog-mass term with a unit winding number, 
\begin{align}
H_{3d}(\bk,\bm{r})
= \bk \cdot \bm{\g}+\bm{r}\cdot \bm{\G}+H_{0d} \G_0, \qquad 
\check g^{(3d)} = \G_0^{\frac{1-c_g}{2}} \check g^{(0d)}, 
\end{align}
with gamma matrices 
\begin{align}
(\bm{\g},\bm{\G},\G_0)
= 
(\s_y s_z\mu_z,s_y\mu_z,\mu_y,\s_x s_z\mu_z,s_x\mu_z,\mu_x,\s_z s_z\mu_z). 
\label{eq:3d_gamma_matrices}
\end{align}
To have the original MPG operators $\hat g$ for the symmetry (\ref{eq:3d_Hamiltonian_mass_hedgehog_sym}), performing the inverse transformation of (\ref{eq:sym_op_ghat_to_gbar}), we have the symmetry operators 
\begin{align}
&\hat g 
= Q_{R_g}^{-1} \times q_{R_g}^{-1} \times (-i \G_3\G_2\G_1\g_3\g_2\g_1)^{\frac{1-s_g}{2}} \times \G_0^{\frac{1-c_g}{2}} \times \check g^{(0d)}. 
\end{align}
This establishes the isomorphism between the group $K_{\rm AI}$ of atomic insulators at the point group center and the group of $3d$ Dirac Hamiltonians with the hedgehog mass with a unit winding number.

\subsection{Homomorphism $f: K_{\rm AI} \to K$}
Following the previous section, we introduce the operator $\tilde g$ acting only on the real space by 
\begin{align}
\tilde g
&:= Q_{R_g}^{-1} \times (-i \G_3\G_2\G_1)^{\frac{1-s_g}{2}} \times \G_0^{\frac{1-c_g}{2}} \times \check g^{(0d)}, 
\label{eq:tildeg_from0d}
\end{align}
\begin{align}
\tilde g H_{3d}(\bk,\bm{r}) \tilde g^{-1}
= s_g c_g H(\phi_g \bk, O_g^{-1} \bm{r}). 
\end{align}
From a straightforward calculation, we see that the factor system of $\tilde g$s matches with $\tilde z_{g,h}$ introduced in (\ref{eq:factor_system_ztilde}). 
Neglecting the hedgehog-mass term $\bm{r} \cdot \bm{\G}$, we have a Dirac Hamiltonian with a uniform mass 
\begin{align}
&H_{3d}'(\bk) := \bm{k} \cdot \bm{\g}+H_{0d}\G_0, \nonumber \\
&\tilde g H_{3d}'(\bk) \tilde g^{-1} 
= s_g c_g H(\phi_g \bk), \qquad \tilde g \tilde h = \tilde z_{g,h} \wt{gh}.
\label{eq:3dHam_from0d}
\end{align}
The Hamiltonian $H_{3d}'(\bk)$ belongs to the $K$-group $K$ and may be reducible. 
Thus, $H_{3d}'(\bk)$ is a direct sum of generators of $K$. 

Computing the homomorphism $f: K_{\rm AI} \to K$ is to establish the character formula for the irreducible decomposition of the $3d$ Dirac Hamiltonians. 
Let $\alpha$ be an irrep of $G_0 = \{g \in G | \phi_g=c_gs_g=1\}$ with the modified factor system $\tilde z_{g,h}$. 
When the EAZ class of $\alpha$ is either of AIII, AIII$_T$, DIII, or CI, the irrep $\tilde \alpha$ contributes to the $K$-group $K$ as the free abelian group $\Z$ characterized by the $3d$ winding number $w_{3d}$. 
When the EAZ class of $\alpha$ is either of AII, AII$_C$ or CII, the irrep $\alpha$ contributes to the $K$-group $K$ as the abelian group $\Z_2$ characterized by the $\Z_2$ number $\nu_{3d}$. 
In the rest of this subsection, we derive the character formulas to give the matrix element $f|_{\beta\to\alpha}: K_{\rm AI}|_{\beta} \to K|_{\alpha}$ of the homomorphism $f$ from a given irrep $\beta$ for $K_{\rm AI}$ to an irrep $\alpha$ for $K$. 

The homomorphism $f: K_{\rm AI} \to K$ takes a form as  $f: \Z^n \times \Z_2^m \to \Z^k \times \Z_2^l$. 
In Appendix~\ref{app:cokernel}, we summarize how to compute the cokernel of $f$ of this type.

\subsubsection{$\Z$ invariant $w_{3d}$ for chiral class}
Given an irrep $\alpha$ of $G_0$, if the chiral operator $ab$ exists and preserves the irrep $\alpha$, namely $O^\G_{\alpha\alpha}=1$, there exists an irrep $\alpha +$ of $G_0 \coprod abG_0=\{g \in G | \phi_g=1\}$ whose restriction on $G_0$ is $\alpha$ (see Appendix~\ref{app:ext_rep}).
By using the irreducible character $\chi_{\alpha +}$ of $\alpha +$, the $3d$ winding number of the Dirac Hamiltonian $H(\bk)=\bk \cdot \bm{\g}+H_{0d}\G_0$ with the symmetry operators $\tilde g$ is given by (see Appendix~\ref{app:3dwinding} for the detail)
\begin{align}
&w_{3d}=(-1) \times \frac{1}{4} \times \frac{1}{|G_0|}\sum_{g \in abG_0}
\chi_{\alpha+}(g)^* \tr 
\left[ 
\g_1\g_2\g_3 
\left\{\begin{array}{ll}
\G_0 & (c_g =1, {}^\forall g \in G)\\
\G_0 \tau_z & (c_g = -1, {}^\exists g \in G)\\
\end{array}\right\} 
\tilde g
\right]. 
\end{align}
The prefactor $(-1)$ was introduced to make the formula simple. 
Plugging the expression (\ref{eq:tildeg_from0d}) of $\tilde g$ into the above formula, we see that for $g \in abG_0$, 
\begin{align}
\g_1\g_2\g_3\G_0\tilde g
&=\g_1\g_2\g_3\G_0 \times Q_{R_g}^{-1} \times (-i \G_3\G_2\G_1)^{\frac{1-s_g}{2}} \times \G_0^{\frac{1-c_g}{2}} \times \check g^{(0d)} \nonumber \\
&=\left\{\begin{array}{ll}
-Q_{R_g}^{-1}\check g^{(0d)} & (c_g=1), \\
\g_1\g_2\g_3Q_{R_g}^{-1}\check g^{(0d)} & (c_g=-1). \\
\end{array}\right.
\end{align}
We have used $\g_1\g_2\g_3\G_0\G_3\G_2\G_1=-i$ for the gamma matrices (\ref{eq:3d_gamma_matrices}). 
Therefore, for the Dirac Hamiltonian (\ref{eq:3dHam_from0d}) obtained by the irrep $\beta$ of $G_0^{\rm AI}$, the $3d$ winding number ${w_{3d}}|_{\beta\to\alpha}$ of the irrep $\alpha$ is 
\begin{align}
{w_{3d}}|_{\beta\to\alpha}
&=(-1) \times \frac{1}{4} \times \frac{1}{|G_0|}\sum_{g \in abG_0} \chi_{\alpha+}(g)^* \delta_{c_g,1} (-1) 
\times 
\left\{\begin{array}{ll}
\tr[Q_{R_g}^{-1} \check g^{(0d)}] & (c_g =1, {}^\forall g \in G),\\
\tr[Q_{R_g}^{-1} \tau_z \check g^{(0d)}] & (c_g = -1, {}^\exists g \in G),\\
\end{array}\right. \nonumber \\
&=\frac{1}{|G_0|}\sum_{g \in G, \atop \phi_g=c_g=-s_g=1} \chi_{\alpha+}(g)^* \times 2 \cos\frac{\theta_g}{2} 
\times 
\left\{\begin{array}{ll}
\tr[\check g^{(0d)}] & (c_g =1, {}^\forall g \in G),\\
\tr[\tau_z \check g^{(0d)}] & (c_g = -1, {}^\exists g \in G), 
\label{eq:cha_formula_w3d}
\\
\end{array}\right.
\end{align}
where we have used $Q_{R_g}^{-1} = \cos \frac{\theta_g}{2} - \sin \frac{\theta_g}{2} \bm{n}_g \cdot \bm{\G}$ and $\tr[\g_1\g_2\g_3Q_{R_g}]=0$. 
The last coefficient is given as 
\begin{align}
&\left\{\begin{array}{ll}
\tr[\check g^{(0d)}] & (c_g =1, {}^\forall g \in G),\\
\tr[\tau_z \check g^{(0d)}] & (c_g = -1, {}^\exists g \in G),\\
\end{array}\right. \nonumber \\
&=\left\{\begin{array}{ll}
\chi_\beta(g) & \mbox{for A, AI}, \\
\chi_\beta(g)+\chi_{\underline{a}(\beta)}(g) & \mbox{for AII, A$_T$}, \\
\chi_\beta(g)-\chi_{\underline{b}(\beta)}(g) & \mbox{for D, C, A$_C$, AI$_C$, BDI, CI}, \\
\chi_\beta(g)-\chi_{\underline{ab}(\beta)}(g) & \mbox{for AIII, A$_\G$}, \\
\chi_\beta(g)+\chi_{\underline{a}(\beta)}(g)-\chi_{\underline{b}(\beta)}(g)-\chi_{\underline{ab}(\beta)}(g) & \mbox{for A$_{T.C}$, AIII$_T$, D$_T$, DIII, AII$_C$, CII, C$_T$}. \\
\end{array}\right.
\end{align}
The matrix element $f|_{\beta \to \alpha}:\Z \to \Z$ is eventually given as 
\begin{align}
f|_{\beta \to \alpha}(1) = 
\left\{\begin{array}{ll}
{w_{3d}}|_{\beta\to\alpha} & (\mbox{for AIII, DIII}), \\ 
\frac{1}{2} \times {w_{3d}}|_{\beta\to\alpha} & (\mbox{for CII}). \\ 
\end{array}\right.
\label{eq:formula_w3d}
\end{align}
Here, the factor $\frac{1}{2}$ for class CII is needed because the $3d$ winding number of class CII takes an even integer. 

\subsubsection{$\Z_2$ invarinat $\nu_{3d}$ for class AII and CII}

For an irrep of $G_0$ whose EAZ class is either of AII, AII$_C$, or CII, the mass term is classified by $\Z_2$. 
Given a Dirac Hamiltonian $H(\bk)=\bk \cdot \bm{\g} + m \G_0$ with the symmetry operator $\tilde g$, the $\Z_2$ invariant $\nu_{3d}$ for the irrep $\alpha$ is given by 
\begin{align}
\nu_{3d}=
\left\{\begin{array}{ll}
\frac{1}{4} \times \#(\mbox{$\alpha$-irreps})\ ({\rm mod\ } 2) & \mbox{for AII, AII$_C$}, \\
\frac{1}{8} \times \#(\mbox{$\alpha$-irreps})\ ({\rm mod\ } 2) & \mbox{for CII}. \\
\end{array}\right.
\end{align}
Here, the number of $\alpha$-irreps, which was denoted by $\#(\mbox{$\alpha$-irreps})$, is given by the irreducible decomposition 
\begin{align}
\#(\mbox{$\alpha$-irreps})
&= 
\frac{1}{|G_0|} \sum_{g \in G_0} \chi_\alpha(g) \tr [\tilde g].
\end{align}
In particular, for the symmetry operators $\tilde g$ given by (\ref{eq:tildeg_from0d}),  noticing $\phi_g=1$ and $c_gs_g=1$ for $g \in G_0$, we have 
\begin{align}
\#(\mbox{$\alpha$-irreps})
&=
\frac{1}{|G_0|} \sum_{g \in G_0} \chi_\alpha(g) \tr \Big[ Q_{R_g}^{-1} \times (-i \G_3\G_2\G_1)^{\frac{1-s_g}{2}} \times \G_0^{\frac{1-c_g}{2}} \times \check g^{(0d)} \Big] \nonumber \\
&= 4 \times \frac{1}{|G_0|} \sum_{g \in G, \atop \phi_g=c_g=s_g=1} \chi_\alpha(g)^* \times 2 \cos \frac{\theta_g}{2} \times \tr[\check g^{(0d)}]. 
\end{align}
Here, we have used $\tr [Q_{R_g}^{-1} \G_3\G_2\G_1\G_0] = 0$.
The last factor is given as
\begin{align}
\tr[\check g^{(0d)}]
&=\left\{\begin{array}{ll}
\chi_\beta(g) & \mbox{for A, AI}, \\
\chi_\beta(g)+\chi_{a(\beta)}(g) & \mbox{for AII, A$_T$}, \\
\chi_\beta(g)+\chi_{b(\beta)}(g) & \mbox{for D, C, A$_C$, AI$_C$, BDI, CI}, \\
\chi_\beta(g)+\chi_{ab(\beta)}(g) & \mbox{for AIII, A$_\G$}, \\
\chi_\beta(g)+\chi_{a(\beta)}(g)+\chi_{b(\beta)}(g)+\chi_{ab(\beta)}(g) & \mbox{for A$_{T.C}$, AIII$_T$, D$_T$, DIII, AII$_C$, CII, C$_T$}. \\
\end{array}\right.
\end{align}
This establishes the formula of the matrix element $f|_{\beta \to \alpha}:\Z \mbox{ or }\Z_2 \to \Z_2$, 
\begin{align}
&f|_{\beta \to \alpha}(1) = {\nu_{3d}}|_{\beta \to \alpha}\nonumber \\
&=\left\{\begin{array}{ll}
1 & \mbox{for AII, AII$_C$}, \\
\frac{1}{2}& \mbox{for CII}. \\
\end{array}\right\}
\times 
\frac{1}{|G_0|} \sum_{g \in G, \atop \phi_g=c_g=p_g=1} \chi_\alpha(g)^* \times 2 \cos \frac{\theta_g}{2} \times \tr[\check g^{(0d)}] \qquad \mbox{mod 2}.
\label{eq:formula_z2}
\end{align}

\section{Classification of surface states of TIs and TSCs}
\label{sec:ti/tscs}
In this section, we apply the irreducible character formulas of the homomorphism $f: K_{\rm AI} \to K$ developed in Sec.~\ref{sec:formulation} to TIs and TSCs with crystallographic MPGs.

The factor system $z_{g,h}$ for MPGs can be arbitrary in general, however, in this paper we concern with spinless or spinful electrons, where the factor system is given by 
\begin{align}
z_{g,h}
=
\left\{\begin{array}{ll}
1 & \mbox{for spinless electrons}, \\
(-1)^{\frac{1-\phi_g}{2} \frac{1-\phi_h}{2}} \times z'_{g,h} & \mbox{for spinful electrons}, \\
\end{array}\right.
\label{eq:factor_system_normal}
\end{align}
with $z'_{g,h} \in \{\pm 1\}$ the factor system of the $Spin(3)$ rotations introduced in (\ref{eq:change_factor_system}). 
The sign $(-1)^{\frac{1-\phi_g}{2} \frac{1-\phi_h}{2}}$ comes from the time-reversal square $\hat T^2=-1$ for spinful electrons.

\subsection{Insulators}
\label{sec:result_ti}
Let $G$ be an MPG. 
Which group element is unitary or antiunitary is specified by the homomorphism $\phi: G \to \{\pm 1\}$. 
For insulators, $c_g$ are identically unity. 
Let $G_0=\{g \in G|\phi_g=s_g=1\}$ be the group of orientation-preserving unitary elements and $T,P$, and $P_t \in G$ be representatives such that 
\begin{align}
-\phi_T=s_T=1, \qquad \phi_P=-s_P=1, \qquad \phi_{P_t}=s_{P_t}=-1, 
\label{eq:TPP_tdef}
\end{align}
respectively. 
The group $G$ splits as 
\begin{align}
G=G_0\coprod TG_0\coprod P_tG_0\coprod PG_0.
\end{align}
For irreps $\alpha$ of $G_0$ with the modified factor system $\tilde z_{g,h}$ defined by (\ref{eq:factor_system_ztilde}), the Winger criteria (\ref{eq:wigner_t}), (\ref{eq:wigner_c}) and the orthogonal test (\ref{eq:ortho}) determine the EAZ classes. 
The results are summarized in Table~\ref{tab:EAZ_TI_spinless} for spinless electrons and Table~\ref{tab:EAZ_TI_spinful} for spinful electrons. 
This extends the previous results for a part of point groups in spinful electrons~\cite{CornfeldPointGroup, OkumaSatoShiozaki}.
From EAZ classes, $K$-groups $K$ are fixed according to the periodic table (Table~\ref{tab:periodictable}). 
For example, the MPG 4/m'mm for spinful electrons has the EAZ classes \{DIII$^2$,AII$_C$\}, meaning that in three space dimensions, the $K$-group, the classification of uniform mass terms, is given by $K=\Z^{\times 2} \times \Z_2$. 

Let $G_0^{\rm AI} = \ker \phi = \{g \in G | \phi_g=1\}$ be the group composed of unitary symmetries. 
The group $G$ splits as 
\begin{align}
G = G_0^{\rm AI} \coprod \underline{a} G_0^{\rm AI}, \qquad \phi_{\underline{a}}=-1, 
\end{align}
with $\underline{a}$ a representative of antiunitary symmetry. 
For an irrep $\beta$ of $G_0^{AI}$ with the factor system $z_{g,h}$, the Wigner criteria (\ref{eq:winger_t_ai}), (\ref{eq:winger_c_ai}) and the orthogonal test (\ref{eq:ortho_ai}) determine the EAZ class of $\beta$ and the $K$-group $K_{\rm AI}$ of atomic insulators at the point group center. 
The formulas (\ref{eq:formula_w3d}) and  (\ref{eq:formula_z2}) give us the homomorphism $f: K_{\rm AI} \to K$. 
The abelian group $K'''$ of 4th-order TIs is given by $K'''= \im f$, and the classification of surface states reads as the quotient group $K/K'''$. 
The groups $K/K'''$ are summarized in Table~\ref{tab:surf_states_ti_spinless} for spinless electrons and Table~\ref{tab:surf_states_ti_spinful} for spinful electrons. 
For example, the surface states compatible with the MPG $\bar1$1' composed of time-reversal and inversion symmetries is classified by $K/K''' = \Z_4$ in spinful electrons~\cite{Haruki230}, where it is known that odd integers of $\Z_4=\{0,1,2,3\}$ correspond to 1st-order TIs, and $2 \in \Z_4$ is the 2nd-order TI that hosts a helical hinge state. 

Some comments are listed below. 
(1) No $1d$ building-block states exist for the class AI and AII. 
Therefore, each element of the group $K/K'''$ represents either of 1st- or 2nd-order TI. 
(2) For MPGs having the ordinary TRS, whose MPG name includes 1', in spinless electrons, we see that no surface states $K/K'''=0$. 
This is compatible with that no building-block states exist for $1d, 2d$ and $3d$ in class AI. 
(3) The 1st-order TI appears if and only if the system is in spinful electrons and the MPG includes the ordinary TRS. 
The ``only if'' part is due to that the $3d$ TI is compatible with $O(3) \times \Z_2^T$ symmetry.~\footnote{
The 1st-order $3d$ TI is represented by the $3d$ Dirac Hamiltonian  
\begin{align*}
    {\cal H}(\bk) = \bm{k} \cdot \bm{\sigma} \tau_z + m \tau_x, 
\end{align*}
where $\bm{\sigma}$ is the spin of electron, and $\tau_\mu$s are the Pauli matrices for the orbital degree of freedom. 
On this basis, the ordinary TRS is given by $T = i \sigma_y {\cal K}$ (${\cal K}$ is the complex conjugation), the $SO(3)$ rotation by $\theta$-angle around the $\bm{n}$-axis is $e^{-i \theta \bm{n} \cdot \bm{\s}/2}$, and the inversion is $I={\rm diag}(-1,-1,-1)$, which constitutes the $O(3)\times \Z_2^T$ symmetry operations. 
}
For other cases, the quotient group $K/K'''$ represents 2nd-order TIs in spinful electrons.

\subsection{Superconductors}
\label{sec:SC}
Let $G$ be an MPG equipped with the homomorphism $\phi_g \in \{\pm 1\}$ and the factor system $z_{g,h}$. 
We assume the normal part $h(\bk)$ of the Bogoliubov-de Gennes (BdG) Hamiltonian is symmetric under the MPG $G$, 
\begin{align}
u_g h(\bk) u_g^{-1}=h(\phi_g O_g \bk), \qquad u_gu_h=z_{g,h}u_{gh}, 
\end{align}
where $u_{g \in G}$ represent the transformations for internal degrees of freedom. 
We assume the superconducting gap function $\Delta(\bk)$ obeys a 1-dimensional representation $e^{i\theta_g} \in U(1)$ of $G$,~\footnote{
When the gap function $\Delta(\bk)$ obeys an  $N$-dimensional irrep $D_\alpha$ of $G$, there is a representation basis $\{\Delta_j(\bk)\}_{j=1}^N$ of the vector space in which the gap function lives satisfying
\begin{align*}
\Delta_i(\phi_g g\bk) = [D_\alpha(g)]_{ij} \times 
\left\{\begin{array}{ll}
u_g \Delta_j(\bk) u_g^T & (\phi_g=1), \\
u_g \Delta_j(\bk)^* u_g^T & (\phi_g=-1). \\
\end{array}\right.
\end{align*}
The gap function $\Delta(\bk)$ is specified by a vector $\bm{\eta} = (\eta_1,\dots,\eta_N) \in \C^N$ so that $\Delta(\bk)=\sum_{j=1}^N \eta_j \Delta_j(\bk)$. 
When the irrep $D_\alpha$ is not a trivial one (namely, an unconventional superconductor), the bare MPG symmetry $G$ is spontaneously broken. 
Using the $U(1)$-phase rotation of complex fermions by the amount of $e^{-i\theta_g/2}$, we can recover the symmetry of $g \in G$ only when $D_\alpha(g)$ is a pure phase $D_\alpha(g)=e^{i\theta_g} \times {\bf 1}$. 
Therefore, we focus only on gap functions obeying 1-dimensional irreps of $G$. 
}
\begin{align}
\Delta(g\bk) = e^{i\theta_g} \times 
\left\{\begin{array}{ll}
u_g \Delta(\bk) u_g^T & (\phi_g=1), \\
u_g \Delta(\bk)^* u_g^T & (\phi_g=-1), \\
\end{array}\right.
\label{eq:Gapfunc_Symm}
\end{align} 
with the trivial factor system $e^{i\theta_g} e^{i \phi_g \theta_h} = e^{i\theta_{gh}}$.~\footnote{
The symmetry constraint (\ref{eq:Gapfunc_Symm}) implies that $e^{i \theta_g} e^{i \phi_g \theta_h} (z_{g,h})^2 = e^{i\theta_{gh}}$. 
Therefore, for spinless electrons ($z_{g,h} \equiv 1$) and spinful electrons ($z_{g,h} \in \{\pm 1\}$), the set of $U(1)$ phases $\{e^{i \theta_g}\}_{g \in G}$ is a 1-dimensional linear representation of $G$. 
}
When $G$ involves an antiunitary element possible 1-dimensional irreps can be listed by the following manner. 
Let $G_u = \ker \phi \subset G$ be the subgroup composed only of unitary elements and $a \in G$ be a representative for antiunitary symmetry such that $G=G_u\coprod aG_u$. 
Given a 1-dimensional irrep $\alpha_{\rm 1d}$ of $G_u$, we have the Wigner criterion $W^T_{\alpha_{\rm 1d}}$ of the irrep $\alpha_{\rm 1d}$ with the trivial factor system. 
If $W^T_{\alpha_{\rm 1d}} \neq 1$, the induced representation of $G$ is not 1-dimensional. 
Thus, $W^T_{\alpha_{\rm 1d}}$ must be $1$. 
Namely, the set of 1-dimensional irreps of $G$ is the set of irreps of $G_u$ with the Wigner criterion $W^T_\alpha=1$. 
For elements $g \in aG_u$, the induced representation of $G$ is given by $e^{-i\theta_{a^{-1}g}}$. 

For the BdG Hamiltonian 
\begin{align}
H(\bk)
=
\begin{pmatrix}
h(\bk)&\Delta(\bk)\\
\Delta(\bk)^\dag&-h(-\bk)^T\\
\end{pmatrix}_\tau, 
\end{align}
the total symmetry group becomes $G \times \Z_2^C$, where $\Z_2^C$ is generated by PHS $\hat C=\tau_x {\cal K}$ with ${\cal K}$ the complex conjugation. 
It should be noted that the symmetry operators $\hat g$ for $g\in G$ depend on the 1-dimensional irrep of the gap function as 
\begin{align}
\hat g = 
\left\{\begin{array}{ll}
\begin{pmatrix}
u_g & \\
& e^{i\theta_g}u_g^* \\
\end{pmatrix} & (\phi_g=1), \\
\begin{pmatrix}
u_g & \\
& e^{i\theta_g}u_g^* \\
\end{pmatrix} \times {\cal K} & (\phi_g=-1). 
\end{array}\right.
\end{align}
Let us fix a factor system for the total symmetry group $G \times \Z_2^C$. 
We define the operators involving $\hat C$ by $\wh{Cg}=\wh{gC}:=\hat C \hat g$ for $g \in G$.
With this choice, by using the equality 
\begin{align}
\hat g \hat C = e^{i\theta_g} \hat C \hat g, \qquad 	g \in G,
\label{eq:gC_vs_Cg}
\end{align}
the factor system is determined as 
\begin{align}
z_{Cg,h} = z_{g,h}^{-1}, \qquad 
z_{g,Ch} = e^{i\theta_g} z_{g,h}^{-1}, \qquad 
z_{Cg,Ch} = e^{-i\theta_g} z_{g,h}, 
\end{align}
for $g,h \in G$. 
In accordance with the general recipe in Sec.~\ref{sec:K}, we introduce the modified operators 
\begin{align}
\tilde g
= 
(\g_1\g_2\g_3)^{\frac{1-s_g}{2}} q_{R_g} \hat g, \qquad g \in G \times \Z_2^C
\end{align}
with the modified factor system 
\begin{align}
\tilde z_{g,h} = (-1)^{\frac{1-c_g\phi_g}{2} \frac{1-s_h}{2}} \times z_{g,h}' \times z_{g,h}, \qquad g,h \in G \times \Z_2^C. 
\end{align}
Note that $z'_{Cg,h}=z'_{g,Ch}=z'_{Cg,Ch}=z'_{g,h}$ for $g,h \in G$. 

It is useful to introduce the the subgroup $G_*:= \{g \in G | \phi_g=s_g=1\} \subset G$ composed of orientation-preserving unitary elements and representatives $T,P$, $P_t\in G$ satisfying equations in (\ref{eq:TPP_tdef}). 
The total symmetry group $G \times \Z_2^C$ is then decomposed as 
\begin{align}
G \times \Z_2^C
&=
\underbrace{(G_*\coprod CP_tG_*)}_{G_0}
\coprod \underbrace{(TG_*\coprod CPG_*)}_{aG_0}
\coprod \underbrace{(P_tG_*\coprod CG_*)}_{bG_0}
\coprod \underbrace{(PG_*\coprod CTG_*)}_{abG_0}.
\end{align}
Given an irrep $\alpha$ of the group $(G_*\coprod CP_tG_*)$ with the factor system $\tilde z_{g,h}$, we get the EAZ class of $\alpha$ from the Winger criteria (\ref{eq:wigner_t}), (\ref{eq:wigner_c}) and the orthogonal test (\ref{eq:ortho}). 
The results are summarized in Table~\ref{tab:EAZ_SC_spinless} for spinless electrons and Table~\ref{tab:EAZ_SC_spinful} for spinful electrons. 

The groups $K_{\rm AI}$, $K/K'''$ are given by the same way as in Sec~\ref{sec:result_ti}. 
Let $G_0^{\rm AI} = \{g \in G | \phi_g=1\} \subset G$ be the subgroup composed of unitary symmetries and $\underline{a} \in G$ be a representative of antiunitary symmetry. 
The total group $G \times \Z_2^C$ splits as 
\begin{align}
G \times \Z_2^C= G_0^{\rm AI} \coprod  \underline{a} G_0^{\rm AI}\coprod CG_0^{\rm AI}\coprod \underline{a}CG_0^{\rm AI}. 
\end{align}
For an irrep $\beta$ of $G_0^{\rm AI}$ with the factor system $z_{g,h}$, the Wigner criteria (\ref{eq:winger_t_ai}), (\ref{eq:winger_c_ai}) and the orthogonal test (\ref{eq:ortho_ai}) determine the EAZ class of $\beta$ and the $K$-group $K_{\rm AI}$. 
The formulas (\ref{eq:formula_w3d}) and  (\ref{eq:formula_z2}) give the homomorphism $f: K_{\rm AI} \to K$. 
The abelian group $K'''$ of 4th-order TSCs is given by $K'''= \im f$, and the classification of surface states is given by the quotient $K/K'''$. 
The results of the quotient groups $K/K'''$ are summarized in Table~\ref{tab:surf_states_sc_spinless} for spinless electrons and Table~\ref{tab:surf_states_sc_spinful} for spinful electrons. 
See Table \ref{tab:1Drep} for the character tables of 1-dimensional irreps of gap functions we used. 

In spinful systems, there may exist the 1st-order TSC if the MPG $G$ includes the ordinary TRS $T$. 
Let us write such MPG by $G=G_{\rm nm} \times \Z_2^T$ with $G_{\rm nm}$ the point group of $G$. 
We first note that the set of 1-dimensional irreps of $G_{\rm nm} \times \Z_2^T$ with the trivial factor system is the set of real 1-dimensional irreps of $G_{\rm nm}$ since $e^{i\theta_g}=e^{i\theta_{TgT}} = e^{i\theta_T} (e^{i\theta_g} e^{i\theta_T})^* = e^{-i\theta_g}$ for $g \in G_{\rm nm}$. 
For the $3d$ Dirac Hamiltonian $H=k_1\g_1+k_2\g_2+k_3\g_3+m\G_0$, the $3d$ winding number detecting the 1st-order TSC is written as 
\begin{align}
w_{3d}^{\rm 1st}=-\frac{1}{4}\tr[\g_1\g_2\g_3\G_0(i \hat C\hat T)].
\end{align}
From equations (\ref{eq:g-sym}), (\ref{eq:gC_vs_Cg}), and $\hat g \hat T=\hat T\hat g$, the winding number $w_{3d}^{\rm 1st}$ changes as 
\begin{align}
w_{3d}^{\rm1st} \xrightarrow[g]{} s_g \times e^{i\theta_g} \times w_{3d}^{\rm 1st} 
\end{align}
under a nonmagnetic point group $g \in G_{\rm nm}$. 
Recall that $s_g=\det[O_g]$ and $e^{i\theta_g} \in \{\pm 1\}$ for $G=G_{\rm nm} \times \Z_2^T$. 
We conclude that 1st-order TSCs survive if and only if the gap function $\Delta(\bk)$ is even under orientation-preserving symmetries and odd under orientation-reversing symmetries for $g \in G_{\rm nm}$. 
In Table~\ref{tab:surf_states_sc_spinful}, the appearance of the 1st-order TSC is highlighted with the bold red character. 

\section{Summary and outlook}
\label{sec:conc}
In this paper, we developed a way to compare the group $K_{\rm AI}$ of atomic insulators with the bulk $K$-group $K$, in the presence MPG symmetry in three space dimensions. 
As an application, we computed the quotient groups $K/K'''$ of the bulk $K$-group $K$ and the group $K'''$ of 4th-order TIs/TSCs for all the 122 MPGs and 1-dimensional representations for the superconducting gap function, which gives the exhaustive classification of surface states of three dimensional TIs and TSCs. 
The main results are summarized in Tables~\ref{tab:EAZ_TI_spinless}-\ref{tab:surf_states_sc_spinful}.

Let us close by mentioning future directions. 

---
The formulation developed in Sec.~\ref{sec:formulation} is applied only to TIs/TSCs without translation invariance. 
To apply our method to magnetic space groups which include a lattice translation, we need to properly glue local building-block Dirac Hamiltonians nearby high-symmetric points together in the whole real space. 
This can be systematically done by the Atiyah-Hirzebruch spectral sequence based on the {\it dual} cell decomposition of the real space, which provides the $E_\infty$-page complementary to the Atiyah-Hirzebruch spectral sequence based on the usual cell decomposition discussed in \cite{KSGeneHomo, HermeleTopologicalCrystal}. 
We leave this problem as future work. 

---
Our formalism can also be applied to the classification of stable superconducting nodal structures in the Brillouin zone. 
For example, a point node at the wave number $\bk_0$ on a high-symmetry line along to the $k_z$-direction is written as a $3d$ gapless Dirac Hamiltonian $H_{\rm pn}(k_x,k_y,k_z)=(\bk-\bk_0) \cdot \bm{\g}$ in the vicinity of $\bk_0$. 
On the one hand, any nodal structures, including point, line, and surface nodes, with the nodal point at $\bk_0$ is described by a gapless $1d$ Dirac Hamiltonian $H_{\rm n}(k_z)=(k_z-k_{z0})\g_z$ on the high-symmetric line. 
Both the types of Hamiltonians $H_{\rm pn}(\bk), H_{\rm n}(k_z)$ are classified and constructed according to the formalism in Sec.~\ref{sec:formulation}. 
Comparing $H_{\rm pn}(\bk)$ and $H_{\rm n}(k_z)$, one can find which a nodal point measured on the high-symmetric line is a point node or not. 

---
In this paper, we focus on 3-space dimensions. 
It should be interesting to generalize our character formulas (\ref{eq:cha_formula_w3d}), (\ref{eq:formula_z2}) to any space dimensions. 


\medskip
\noindent 
{\it Acknowledgement---}
I thank Luka Trifunovic for a useful discussions. 
I am grateful to Eyal Cornfeld for teaching me the relationship between $SO(3)$ rotation and quaternion. 
This work was supported by PRESTO, JST (Grant No. JPMJPR18L4).

\appendix

\section{An extension of irreducible representations}
\label{app:ext_rep}

Let $G$ be a finite group equipped with the factor system $\hat g \hat h = z_{g,h} \wh{gh}$. 
Suppose the group $G$ sits into the short exact sequence $G_0 \to G \to \{e,ab\}$.
For an irrep $\alpha$ of $G_0$, if the mapped representation $ab(\alpha)$ is unitary equivalent to $\alpha$ or not is determined by the orthogonal test 
\begin{align}
&O_{\alpha\alpha}^{\G}
= 
\frac{1}{|G_0|} \sum_{g \in G_0} \left[ \frac{z_{g,ab}}{z_{ab,(ab)^{-1}g ab}} \chi_{\alpha}((ab)^{-1}g ab) \right]^* \chi_{\alpha}(g) \in \{0,1\}.
\end{align}
Here $\chi_\alpha(g \in G_0)$ is the irreducible character of $\alpha$.  
Let $D_\alpha(g)$ be a matrix representation of $\alpha$. 
When $O^\G_{\alpha\alpha}=1$, there exists a unitary matrix $U$ such that the relations 
\begin{align}
\frac{z_{g,ab}}{z_{ab,(ab)^{-1}gab}} D_\alpha((ab)^{-1}gab)
=
U^\dag D_\alpha(g) U, \qquad g \in G_0, 
\end{align}
hold true. 
From a straightforward calculation, we find that $[U,D_\alpha((ab)^2)]=0$ and $[D_\alpha(g),D_\alpha((ab)^2)U^{-2}]=0$ for $g \in G_0$. 
The latter implies, from the Schur's lemma, $U^2 = \lambda D_\alpha((ab)^2)$ with a $U(1)$ phase $\lambda$. 
We fix the $U(1)$ phase by demanding $U^2 = z_{ab,ab}D_\alpha((ab)^2)$, that is, $\lambda = \pm \sqrt{z_{ab,ab}}$. 
Picking up a sign of the square root, we set $D_{\alpha+}(ab):=U$. 
For the representation matrices for other elements $g \in ab G_0$, we define $D_{\alpha+}(g(ab)) :=z_{g,ab}^{-1} D_\alpha(g) U =z_{ab,(ab)^{-1}g(ab)}^{-1} U D_\alpha((ab)^{-1}g(ab))$ for $g \in G_0$. 
One can show that the set of matrices $\{ D_{\alpha+}(g) \}_{g \in G_0\coprod abG_0}$ obeys a projective representation of $G_0 \coprod abG_0$ with the factor system $z_{g,h}$. 
We note that $D_{\alpha+}$ should be irreducible as a representation of $G_0 \coprod abG_0$, because the restricted one $D_{\alpha}$ on $G_0$ of $D_{\alpha+}$ is irreducible. 

The alternative choice $D_{\alpha-}(ab):=-U$ gives another  inequivalent irrep of $G_0 \coprod ab G_0$. 
In fact, using the equality 
\begin{align}
D_{\alpha-}(g)=\left\{\begin{array}{ll}
D_{\alpha+}(g) & (g \in G_0), \\
-D_{\alpha+}(g) & (g \in abG_0), \\
\end{array}\right.
\end{align}
we have 
\begin{align}
&\frac{1}{|G_0\coprod abG_0|} \sum_{g \in (G_0\coprod abG_0)} [\tilde D_{\alpha+}(g)^*]_{ij} [\tilde D_{\alpha-}(g)]_{kl} \nonumber \\
&=\frac{1}{2|G_0|}
\sum_{g \in G_0} \left[
D_{\alpha+}(g)^*]_{ij} [ D_{\alpha+}(g)]_{kl} - [ D_{\alpha+}(gs)^*]_{ij} [ D_{\alpha+}(gs)]_{kl}
\right] \nonumber \\
&=\frac{1}{2|G_0|} 
\sum_{g \in G_0} \left[
[D_\alpha(g)^*]_{ij} [D_\alpha(g)]_{kl} - [D_\alpha(g)^*]_{im} U^*_{mj} [D_\alpha(g)]_{kn} U_{nl}
\right]\nonumber \\
&=0. 
\end{align}
Here, we used the orthogonality relation among irreps 
\begin{align}
\frac{1}{|G_0|} \sum_{g \in G_0} [D_\alpha(g)^*]_{ij} [D_\beta(g)]_{kl} 
=\frac{1}{{\rm dim}(\alpha)} \delta_{\alpha\beta} \delta_{ik} \delta_{jl}. 
\end{align}

\section{The $3d$ winding number for Dirac Hamiltonians}
\label{app:3dwinding}
Before going on constructing the $3d$ winding number $w_{3d}$ for generic cases, we first consider the cases where the chiral symmetry $\G$ is the only symmetry of the system, i.e., AIII of AZ class. 
For the Dirac Hamiltonian $H(\bk) = \bk \cdot \bm{\g} + m\G_0$ with chiral symmetry $\G H(\bk) \G^{-1}=-H(\bk)$, the $3d$ winding number $w_{3d}$ is given by 
\begin{align}
w_{3d}=-\frac{1}{4} \tr[\g_1\g_2\g_3\G_0\G].
\label{app:3d_winding_number_aiii}
\end{align}
Actually, the minimal 4 by 4 model $H(\bk)=\bk \cdot \bm{\s}\mu_x+m\mu_y, \G=\mu_z$ has $w_{3d}=1$. 
We should suitably generalize this formula to generic cases. 

Let us consider $3d$ Dirac Hamiltonian $H(\bk)=\bk\cdot \bm{\g}+m\G_0$ with unitary antisymmetry 
\begin{align}
\tilde g H(\bk) \tilde g^{-1}=c_gs_gH(\bk), \qquad g \in G_0\coprod abG_0. 
\end{align}
Let $\alpha$ be an irrep with the orthogonal test $O^\G_{\alpha\alpha}=1$ so that the $3d$ winding number $w_{3d}$ is well-defined. 
The integer $w_{3d}$ counts how many times the "irreducible" $3d$ Dirac Hamiltonians made from the irrep $\alpha$ occur in $H(\bk)$. 
A subtle point is that the $\tilde g$ square for $g \in abG_0$ is not proportional to the identity operator in general, which spoils the formula (\ref{app:3d_winding_number_aiii}). 
Instead, we employ the extended irreps $D_{\alpha+}$ and $D_{\alpha-}$ introduced in Appendix~\ref{app:ext_rep}. 
The irreps $D_{\alpha+}$ and $D_{\alpha-}$ play the roles of the positive and negative chiralities. 
To apply the orthogonality relation of the irreducible character, we introduce new operators $\rho(g)$ for $g \in G_0 \coprod abG_0$ with the same factor system $\tilde z_{g,h}$ as 
\begin{align}
\rho(g):= (-\g_1\g_2\g_3\G_0)^{\frac{1-c_gs_g}{2}} \tilde g, \qquad g \in G_0\coprod abG_0, \qquad 
\qquad \rho(g)\rho(h)=\tilde z_{g,h}\rho(gh), 
\end{align}
so that $\rho(g)$ behaves unitary symmetry 
\begin{align}
\rho(g) H(\bk) \rho(g)^{-1}=H(\bk), \qquad 
g \in G_0\coprod abG_0.
\end{align}
Then, given a $3d$ Dirac Hamiltonian $H(\bk)$ and symmetry operators $\tilde g$, $w_{3d}$ is given as 
\begin{align}
w_{3d}
&= \frac{1}{4} \left[ 
\#(\mbox{$\alpha+$-irreps in $\rho$}) - \#(\mbox{$\alpha-$-irreps in $\rho$})
\right]\nonumber \\
&= \frac{1}{4} \times \frac{1}{|G_0\coprod abG_0|} \sum_{g \in G_0\coprod abG_0}
\left\{\chi_{\alpha+}(g)-\chi_{\alpha-}(g) \right\}^*
\times \tr[\rho(g)]
\nonumber \\
&= -\frac{1}{4} \times \frac{1}{|G_0|} \sum_{g \in abG_0}
\chi_{\alpha+}(g)^*
\times \tr[\g_1\g_2\g_3\G_0\tilde g].
\end{align}

\section{The cokernel of $f: K_{\rm AI} \to K$}
\label{app:cokernel}
In this Appendix, we formulate how to compute the cokernel of the homomorphism between abelian groups involving $\Z$ and $\Z_2$. 
Let us consider a homomorphism 
\begin{align}
f: \Z^{\oplus n} \oplus \Z_2^{\oplus m} \to \Z^{\oplus k} \oplus \Z_2^{\oplus l}.
\end{align}
Let $\{x_j\}_{j=1}^n, \{y_j\}_{j=1}^m, \{z_j\}_{j=1}^k, \{w_j\}_{j=1}^l$ be bases of $\Z^{\oplus n}$, $\Z_2^{\oplus m}$, $\Z^{\oplus k}$, $\Z_2^{\oplus l}$, respectively. 
The representative matrix $M$, which is defined by 
\begin{align}
f(x_1,\dots,x_n;y_1,\dots,y_m) = (z_1,\dots,z_k;w_1,\dots,w_l) M, 
\end{align}
is written as 
\begin{align}
M = 
\left[
\begin{array}{c|ccccccccc}
A & O \\
\hline 
B & C \\
\end{array}
\right], \qquad 
A \in {\rm Mat}_{k \times n}(\Z), \qquad 
B \in {\rm Mat}_{l \times n}(\Z_2), \qquad 
C \in {\rm Mat}_{l \times m}(\Z_2). 
\end{align}
Applying the Smith decomposition to $C$, we have 
\begin{align}
u C v= 
\left[
\begin{array}{c|ccccccccc}
{\bf 1}_p & O \\
\hline 
O & O \\
\end{array}
\right]
\end{align}
with $u,v$ unimodular matrices. 
Then, $M$ is written as 
\begin{align}
M
&=
\left[
\begin{array}{c|c}
{\bf 1}_n & O\\
\hline
O& u^{-1} \\
\end{array}
\right]
\left[
\begin{array}{c|c}
A&O\\
\hline
uB&
\begin{array}{c|c}
{\bf 1}_p&O\\
\hline
O&O\\
\end{array}
\\
\end{array}
\right]
\left[
\begin{array}{c|c}
{\bf 1}_n & O\\
\hline
O& v^{-1} \\
\end{array}
\right]\nonumber \\
&=
\left[
\begin{array}{c|c}
{\bf 1}_n & O\\
\hline
O& u^{-1} \\
\end{array}
\right]
\left[
\begin{array}{c|c}
A&O\\
\hline
\begin{array}{c}
O \\
\hline
{[}uB]_{\rm even} \\
\end{array}
&
\begin{array}{c|c}
{\bf 1}_p&O\\
\hline
O&O\\
\end{array}
\\
\end{array}
\right]
\left[
\begin{array}{c|c}
{\bf 1}_n & O\\
\hline
\begin{array}{c}
{[}uB]_{\rm odd}\\
\hline
O
\end{array}
& v^{-1} \\
\end{array}
\right].  
\end{align}
Here, we have introduced the notation of submatrices 
\begin{align}
uB = 
\left[
\begin{array}{ccccccc}
{[}uB]_{1,1} & \cdots & [uB]_{1,n}\\
\vdots&&\vdots\\
{[}uB]_{p,1} & \cdots & [uB]_{p,n}\\
\hline
{[}uB]_{p+1,1} & \cdots & [uB]_{p+1,n}\\
\vdots&&\vdots\\
{[}uB]_{l,1} & \cdots & [uB]_{l,n}\\
\end{array}
\right]
=: 
\left[
\begin{array}{c}
{[}uB]_{\rm odd} \\
\hline
{[}uB]_{\rm even} \\
\end{array}
\right]
\end{align}
The problem to compute the cokernel of $f$ is recast as that of the restricted homomorphism 
\begin{align}
f': \Z^{\oplus n} \to \Z^{\oplus m} \oplus \Z_2^{\oplus (l-p)} 
\end{align}
with the representation matrix 
\begin{align}
M'
= 
\left[
\begin{array}{c}
A \\
{[}uB]_{\rm even}\\
\end{array}
\right].
\end{align}
This can be done by embedding $\Z_2$ into $\Z$, and taking the quotient of the homomorphism $\Z \xrightarrow{2} \Z$. 
The cokernel of $f'$ is the same as that of the homomorphism $f'': \Z^{\oplus (n+l-p)} \to \Z^{\oplus(m+l-p)}$ with the representation matrix 
\begin{align}
M''=
\left[
\begin{array}{cc}
A&O\\
{[}uB]_{\rm even}&{\bf 2}_{l-p}\\
\end{array}
\right].
\end{align}
Applying the Smith decomposition to $M''$, we have 
\begin{align}
u'M''v' = 
\left[
\begin{array}{c|c}
\begin{array}{cccc}
d_1\\
&d_2\\
&&\ddots\\
&&&d_q\\
\end{array}&O\\
\hline 
O_{m+l-p-q,q}&O\\
\end{array}
\right]
\end{align}
with $d_j (j=1,\dots,q)$ nonnegative integers. 
The cokernel of $f$ is eventually given by 
\begin{align}
{\rm coker} f
\cong 
\Z^{\oplus (m+l-p-q)} \oplus \bigoplus_{j=1}^q \Z_{d_j}. 
\end{align}

\section{Classification tables}
\label{app:tables}
This appendix summarizes the classification tables. 
In Tables~\ref{tab:surf_states_ti_spinless}, \ref{tab:surf_states_ti_spinful}, \ref{tab:surf_states_sc_spinless}, and  \ref{tab:surf_states_sc_spinful}, ``Free'' and ``Tor'' stand for the free and torsion parts of the quotient group $K/K'''$, respectively.

\begin{table}
\caption{EAZ classes for spinless electrons with crystallographic MPG symmetry.}
\label{tab:EAZ_TI_spinless}
\begin{align*}
\begin{array}{ll}
\mbox{MPG}&\mbox{EAZ}\\
\hline 
 1 & \{\text{A}\} \\
 \text{11'} & \{\text{AI}\} \\
 \bar{1} & \{\text{AIII}\} \\
 \bar{1}\text{1'} & \{\text{CI}\} \\
 \bar{1}' & \{\text{C}\} \\
 2 & \left\{\text{A}^2\right\} \\
 \text{21'} & \{\text{A}_T\} \\
 \text{2'} & \{\text{AII}\} \\
 \text{m} & \{\text{AIII}\} \\
 \text{m1'} & \{\text{BDI}\} \\
 \text{m'} & \{\text{D}\} \\
 \text{2/m} & \left\{\text{AIII}^2\right\} \\
 \text{2/m1'} & \{\text{AIII}_T\} \\
 \text{2'/m} & \{\text{CII}\} \\
 \text{2/m'} & \{\text{A}_C\} \\
 \text{2'/m'} & \{\text{DIII}\} \\
 222 & \{\text{A}\} \\
 \text{2221'} & \{\text{AII}\} \\
 \text{2'2'2} & \left\{\text{AII}^2\right\} \\
 \text{mm2} & \{\text{A}_\Gamma\} \\
 \text{mm21'} & \{\text{D}_T\} \\
 \text{m'm2'} & \{\text{DIII}\} \\
 \text{m'm'2} & \left\{\text{D}^2\right\} \\
 \text{mmm} & \{\text{AIII}\} \\
 \text{mmm1'} & \{\text{DIII}\} \\
 \text{m'mm} & \{\text{AII}_C\} \\
 \text{m'm'm} & \left\{\text{DIII}^2\right\} \\
 \text{m'm'm'} & \{\text{D}\} \\
 4 & \left\{\text{A}^4\right\} \\
 \text{41'} & \left\{\text{A}_T{}^2\right\} \\
 \text{4'} & \{\text{A}_T\} \\
 \bar{4} & \left\{\text{AIII}^2\right\} \\
 \bar{4}\text{1'} & \{\text{AIII}_T\} \\
 \bar{4}' & \{\text{A}_C\} \\
 \text{4/m} & \left\{\text{AIII}^4\right\} \\
 \text{4/m1'} & \left\{\text{AIII}_T{}^2\right\} \\
 \text{4'/m} & \{\text{AIII}_T\} \\
 \text{4/m'} & \left\{\text{A}_C{}^2\right\} \\
 \text{4'/m'} & \{\text{AIII}_T\} \\
 422 & \left\{\text{A}^2\right\} \\
 \text{4221'} & \left\{\text{AII}^2\right\} \\
\end{array}&&
\begin{array}{ll}
\mbox{MPG}&\mbox{EAZ}\\
\hline 
 \text{4'22'} & \{\text{AII}\} \\
 \text{42'2'} & \left\{\text{AII}^4\right\} \\
 \text{4mm} & \left\{\text{A}_\Gamma{}^2\right\} \\
 \text{4mm1'} & \left\{\text{D}_T{}^2\right\} \\
 \text{4'm'm} & \{\text{D}_T\} \\
 \text{4m'm'} & \left\{\text{D}^4\right\} \\
 \bar{4}\text{2m} & \{\text{AIII}\} \\
 \bar{4}\text{2m1'} & \{\text{DIII}\} \\
 \bar{4}\text{'2'm} & \{\text{AII}_C\} \\
 \bar{4}\text{'2m'} & \{\text{D}\} \\
 \bar{4}\text{2'm'} & \left\{\text{DIII}^2\right\} \\
 \text{4/mmm} & \left\{\text{AIII}^2\right\} \\
 \text{4/mmm1'} & \left\{\text{DIII}^2\right\} \\
 \text{4/m'mm} & \left\{\text{AII}_C{}^2\right\} \\
 \text{4'/mm'm} & \{\text{DIII}\} \\
 \text{4'/m'm'm} & \{\text{DIII}\} \\
 \text{4/mm'm'} & \left\{\text{DIII}^4\right\} \\
 \text{4/m'm'm'} & \left\{\text{D}^2\right\} \\
 3 & \left\{\text{A}^3\right\} \\
 \text{31'} & \{\text{AI},\text{A}_T\} \\
 \bar{3} & \left\{\text{AIII}^3\right\} \\
 \bar{3}\text{1'} & \{\text{AIII}_T,\text{CI}\} \\
 \bar{3}' & \{\text{C},\text{A}_C\} \\
 32 & \left\{\text{A}^3\right\} \\
 \text{321'} & \{\text{AII},\text{A}_T\} \\
 \text{32'} & \left\{\text{AII}^3\right\} \\
 \text{3m} & \{\text{AIII},\text{A}_\Gamma\} \\
 \text{3m1'} & \{\text{BDI},\text{D}_T\} \\
 \text{3m'} & \left\{\text{D}^3\right\} \\
 \bar{3}\text{m} & \left\{\text{AIII}^3\right\} \\
 \bar{3}\text{m1'} & \{\text{AIII}_T,\text{DIII}\} \\
 \bar{3}\text{'m} & \{\text{AII}_C,\text{CII}\} \\
 \bar{3}\text{'m'} & \{\text{D},\text{A}_C\} \\
 \bar{3}\text{m'} & \left\{\text{DIII}^3\right\} \\
 6 & \left\{\text{A}^6\right\} \\
 \text{61'} & \left\{\text{A}_T{}^3\right\} \\
 \text{6'} & \{\text{AII},\text{A}_T\} \\
 \bar{6} & \left\{\text{AIII}^3\right\} \\
 \bar{6}\text{1'} & \{\text{AIII}_T,\text{BDI}\} \\
 \bar{6}' & \{\text{D},\text{A}_C\} \\
 \text{6/m} & \left\{\text{AIII}^6\right\} \\
\end{array}&&
\begin{array}{ll}
\mbox{MPG}&\mbox{EAZ}\\
\hline 
 \text{6/m1'} & \left\{\text{AIII}_T{}^3\right\} \\
 \text{6'/m} & \{\text{AIII}_T,\text{CII}\} \\
 \text{6/m'} & \left\{\text{A}_C{}^3\right\} \\
 \text{6'/m'} & \{\text{AIII}_T,\text{DIII}\} \\
 622 & \left\{\text{A}^3\right\} \\
 \text{6221'} & \left\{\text{AII}^3\right\} \\
 \text{6'22'} & \left\{\text{AII}^3\right\} \\
\end{array}
\end{align*}
\end{table}

\begin{table}
\caption{EAZ classes for spinful electrons with crystallographic MPG symmetry.}
\label{tab:EAZ_TI_spinful}
\begin{align*}
\begin{array}{ll}
\mbox{MPG}&\mbox{EAZ}\\
\hline 
  1 & \{\text{A}\} \\
 \text{11'} & \{\text{AII}\} \\
 \bar{1} & \{\text{AIII}\} \\
 \bar{1}\text{1'} & \{\text{DIII}\} \\
 \bar{1}' & \{\text{D}\} \\
 2 & \left\{\text{A}^2\right\} \\
 \text{21'} & \left\{\text{AII}^2\right\} \\
 \text{2'} & \{\text{AII}\} \\
 \text{m} & \{\text{AIII}\} \\
 \text{m1'} & \{\text{DIII}\} \\
 \text{m'} & \{\text{D}\} \\
 \text{2/m} & \left\{\text{AIII}^2\right\} \\
 \text{2/m1'} & \left\{\text{DIII}^2\right\} \\
 \text{2'/m} & \{\text{DIII}\} \\
 \text{2/m'} & \left\{\text{D}^2\right\} \\
 \text{2'/m'} & \{\text{DIII}\} \\
 222 & \left\{\text{A}^4\right\} \\
 \text{2221'} & \left\{\text{AII}^4\right\} \\
 \text{2'2'2} & \left\{\text{AII}^2\right\} \\
 \text{mm2} & \left\{\text{AIII}^2\right\} \\
 \text{mm21'} & \left\{\text{DIII}^2\right\} \\
 \text{m'm2'} & \{\text{DIII}\} \\
 \text{m'm'2} & \left\{\text{D}^2\right\} \\
 \text{mmm} & \left\{\text{AIII}^4\right\} \\
 \text{mmm1'} & \left\{\text{DIII}^4\right\} \\
 \text{m'mm} & \left\{\text{DIII}^2\right\} \\
 \text{m'm'm} & \left\{\text{DIII}^2\right\} \\
 \text{m'm'm'} & \left\{\text{D}^4\right\} \\
 4 & \left\{\text{A}^4\right\} \\
 \text{41'} & \left\{\text{AII}^2,\text{A}_T\right\} \\
 \text{4'} & \{\text{AI},\text{AII}\} \\
 \bar{4} & \left\{\text{AIII}^2\right\} \\
 \bar{4}\text{1'} & \{\text{DIII},\text{CII}\} \\
 \bar{4}' & \{\text{D},\text{C}\} \\
 \text{4/m} & \left\{\text{AIII}^4\right\} \\
 \text{4/m1'} & \left\{\text{AIII}_T,\text{DIII}^2\right\} \\
 \text{4'/m} & \{\text{DIII},\text{CI}\} \\
 \text{4/m'} & \left\{\text{D}^2,\text{A}_C\right\} \\
 \text{4'/m'} & \{\text{BDI},\text{DIII}\} \\
 422 & \left\{\text{A}^5\right\} \\
 \text{4221'} & \left\{\text{AII}^5\right\} \\
\end{array}&&
\begin{array}{ll}
\mbox{MPG}&\mbox{EAZ}\\
\hline 
 \text{4'22'} & \left\{\text{AII}^2,\text{A}_T\right\} \\
 \text{42'2'} & \left\{\text{AII}^4\right\} \\
 \text{4mm} & \left\{\text{AIII}^2,\text{A}_\Gamma\right\} \\
 \text{4mm1'} & \left\{\text{D}_T,\text{DIII}^2\right\} \\
 \text{4'm'm} & \{\text{BDI},\text{DIII}\} \\
 \text{4m'm'} & \left\{\text{D}^4\right\} \\
 \bar{4}\text{2m} & \left\{\text{AIII}^2,\text{A}_\Gamma\right\} \\
\end{array}&&
\begin{array}{ll}
\mbox{MPG}&\mbox{EAZ}\\
\hline 
 \text{6/m1'} & \left\{\text{AIII}_T{}^2,\text{DIII}^2\right\} \\
 \text{6'/m} & \{\text{AIII}_T,\text{DIII}\} \\
 \text{6/m'} & \left\{\text{D}^2,\text{A}_C{}^2\right\} \\
 \text{6'/m'} & \{\text{AIII}_T,\text{DIII}\} \\
 622 & \left\{\text{A}^6\right\} \\
 \text{6221'} & \left\{\text{AII}^6\right\} \\
 \text{6'22'} & \left\{\text{AII}^3\right\} \\
\end{array}
\end{align*}
\end{table}

\begin{table}
\caption{
The classification of surface states of $3d$ TIs of spinless electrons with crystallographic MPG symmetry.
}
\label{tab:surf_states_ti_spinless}
\begin{align*}
\begin{array}{lll}
\mbox{MPG}&\mbox{Free}&\mbox{Tor}\\
\hline 
 1 & 0 & \{\} \\
 \text{11'} & 0 & \{\} \\
 \bar{1} & 0 & \{2\} \\
 \bar{1}\text{1'} & 0 & \{\} \\
 \bar{1}' & 0 & \{\} \\
 2 & 0 & \{\} \\
 \text{21'} & 0 & \{\} \\
 \text{2'} & 0 & \{2\} \\
 \text{m} & 1 & \{\} \\
 \text{m1'} & 0 & \{\} \\
 \text{m'} & 0 & \{\} \\
 \text{2/m} & 1 & \{\} \\
 \text{2/m1'} & 0 & \{\} \\
 \text{2'/m} & 0 & \{\} \\
 \text{2/m'} & 0 & \{\} \\
 \text{2'/m'} & 0 & \{2\} \\
 222 & 0 & \{\} \\
 \text{2221'} & 0 & \{\} \\
 \text{2'2'2} & 0 & \{2\} \\
 \text{mm2} & 0 & \{\} \\
 \text{mm21'} & 0 & \{\} \\
 \text{m'm2'} & 1 & \{\} \\
 \text{m'm'2} & 0 & \{\} \\
 \text{mmm} & 0 & \{\} \\
 \text{mmm1'} & 0 & \{\} \\
 \text{m'mm} & 0 & \{\} \\
 \text{m'm'm} & 1 & \{\} \\
 \text{m'm'm'} & 0 & \{\} \\
 4 & 0 & \{\} \\
 \text{41'} & 0 & \{\} \\
 \text{4'} & 0 & \{\} \\
 \bar{4} & 0 & \{2\} \\
 \bar{4}\text{1'} & 0 & \{\} \\
 \bar{4}' & 0 & \{\} \\
 \text{4/m} & 1 & \{\} \\
 \text{4/m1'} & 0 & \{\} \\
 \text{4'/m} & 0 & \{\} \\
 \text{4/m'} & 0 & \{\} \\
 \text{4'/m'} & 0 & \{\} \\
 422 & 0 & \{\} \\
 \text{4221'} & 0 & \{\} \\
\end{array}&&
\begin{array}{lll}
\mbox{MPG}&\mbox{Free}&\mbox{Tor}\\
\hline 
 \text{4'22'} & 0 & \{\} \\
 \text{42'2'} & 0 & \{2\} \\
 \text{4mm} & 0 & \{\} \\
 \text{4mm1'} & 0 & \{\} \\
 \text{4'm'm} & 0 & \{\} \\
 \text{4m'm'} & 0 & \{\} \\
 \bar{4}\text{2m} & 0 & \{\} \\
 \bar{4}\text{2m1'} & 0 & \{\} \\
 \bar{4}\text{'2'm} & 0 & \{\} \\
 \bar{4}\text{'2m'} & 0 & \{\} \\
 \bar{4}\text{2'm'} & 0 & \{2\} \\
 \text{4/mmm} & 0 & \{\} \\
 \text{4/mmm1'} & 0 & \{\} \\
 \text{4/m'mm} & 0 & \{\} \\
 \text{4'/mm'm} & 0 & \{\} \\
 \text{4'/m'm'm} & 0 & \{\} \\
 \text{4/mm'm'} & 1 & \{\} \\
 \text{4/m'm'm'} & 0 & \{\} \\
 3 & 0 & \{\} \\
 \text{31'} & 0 & \{\} \\
 \bar{3} & 0 & \{2\} \\
 \bar{3}\text{1'} & 0 & \{\} \\
 \bar{3}' & 0 & \{\} \\
 32 & 0 & \{\} \\
 \text{321'} & 0 & \{\} \\
 \text{32'} & 0 & \{2\} \\
 \text{3m} & 1 & \{\} \\
 \text{3m1'} & 0 & \{\} \\
 \text{3m'} & 0 & \{\} \\
 \bar{3}\text{m} & 1 & \{\} \\
 \bar{3}\text{m1'} & 0 & \{\} \\
 \bar{3}\text{'m} & 0 & \{\} \\
 \bar{3}\text{'m'} & 0 & \{\} \\
 \bar{3}\text{m'} & 0 & \{2\} \\
 6 & 0 & \{\} \\
 \text{61'} & 0 & \{\} \\
 \text{6'} & 0 & \{2\} \\
 \bar{6} & 1 & \{\} \\
 \bar{6}\text{1'} & 0 & \{\} \\
 \bar{6}' & 0 & \{\} \\
 \text{6/m} & 1 & \{\} \\
\end{array}&&
\begin{array}{lll}
\mbox{MPG}&\mbox{Free}&\mbox{Tor}\\
\hline 
 \text{6/m1'} & 0 & \{\} \\
 \text{6'/m} & 0 & \{\} \\
 \text{6/m'} & 0 & \{\} \\
 \text{6'/m'} & 0 & \{2\} \\
 622 & 0 & \{\} \\
 \text{6221'} & 0 & \{\} \\
 \text{6'22'} & 0 & \{2\} \\
 \text{62'2'} & 0 & \{2\} \\
 \text{6mm} & 0 & \{\} \\
 \text{6mm1'} & 0 & \{\} \\
 \text{6'mm'} & 1 & \{\} \\
 \text{6m'm'} & 0 & \{\} \\
 \bar{6}\text{m2} & 0 & \{\} \\
 \bar{6}\text{m21'} & 0 & \{\} \\
 \bar{6}\text{'m'2} & 0 & \{\} \\
 \bar{6}\text{'m2'} & 1 & \{\} \\
 \bar{6}\text{m'2'} & 1 & \{\} \\
 \text{6/mmm} & 0 & \{\} \\
 \text{6/mmm1'} & 0 & \{\} \\
 \text{6/m'mm} & 0 & \{\} \\
 \text{6'/mmm'} & 0 & \{\} \\
 \text{6'/m'mm'} & 1 & \{\} \\
 \text{6/mm'm'} & 1 & \{\} \\
 \text{6/m'm'm'} & 0 & \{\} \\
 23 & 0 & \{\} \\
 \text{231'} & 0 & \{\} \\
 \text{m}\bar{3} & 0 & \{\} \\
 \text{m}\bar{3}\text{1'} & 0 & \{\} \\
 \text{m'}\bar{3}' & 0 & \{\} \\
 432 & 0 & \{\} \\
 \text{4321'} & 0 & \{\} \\
 \text{4'32'} & 0 & \{\} \\
 \bar{4}\text{3m} & 0 & \{\} \\
 \bar{4}\text{3m1'} & 0 & \{\} \\
 \bar{4}\text{'3m'} & 0 & \{\} \\
 \text{m}\bar{3}\text{m} & 0 & \{\} \\
 \text{m}\bar{3}\text{m1'} & 0 & \{\} \\
 \text{m'}\bar{3}\text{'m} & 0 & \{\} \\
 \text{m}\bar{3}\text{m'} & 0 & \{\} \\
 \text{m'}\bar{3}\text{'m'} & 0 & \{\} \\
\ \\
\end{array}
\end{align*}
\end{table}

\begin{table}
\caption{The classification of surface states of $3d$ TIs of spinful electrons with crystallographic MPG symmetry.}
\label{tab:surf_states_ti_spinful}
\begin{align*}
\begin{array}{lll}
\mbox{MPG}&\mbox{Free}&\mbox{Tor}\\
\hline 
 1 & 0 & \{\} \\
 \text{11'} & 0 & \{2\} \\
 \bar{1} & 0 & \{2\} \\
 \bar{1}\text{1'} & 0 & \{4\} \\
 \bar{1}' & 0 & \{\} \\
 2 & 0 & \{\} \\
 \text{21'} & 0 & \{2,2\} \\
 \text{2'} & 0 & \{2\} \\
 \text{m} & 1 & \{\} \\
 \text{m1'} & 1 & \{\} \\
 \text{m'} & 0 & \{\} \\
 \text{2/m} & 1 & \{\} \\
 \text{2/m1'} & 1 & \{2\} \\
 \text{2'/m} & 1 & \{\} \\
 \text{2/m'} & 0 & \{\} \\
 \text{2'/m'} & 0 & \{2\} \\
 222 & 0 & \{\} \\
 \text{2221'} & 0 & \{2,2,2\} \\
 \text{2'2'2} & 0 & \{2\} \\
 \text{mm2} & 2 & \{\} \\
 \text{mm21'} & 2 & \{\} \\
 \text{m'm2'} & 1 & \{\} \\
 \text{m'm'2} & 0 & \{\} \\
 \text{mmm} & 3 & \{\} \\
 \text{mmm1'} & 3 & \{\} \\
 \text{m'mm} & 2 & \{\} \\
 \text{m'm'm} & 1 & \{\} \\
 \text{m'm'm'} & 0 & \{\} \\
 4 & 0 & \{\} \\
 \text{41'} & 0 & \{2,2\} \\
 \text{4'} & 0 & \{2\} \\
 \bar{4} & 0 & \{2\} \\
 \bar{4}\text{1'} & 0 & \{4\} \\
 \bar{4}' & 0 & \{\} \\
 \text{4/m} & 1 & \{\} \\
 \text{4/m1'} & 1 & \{2\} \\
 \text{4'/m} & 1 & \{\} \\
 \text{4/m'} & 0 & \{\} \\
 \text{4'/m'} & 0 & \{2\} \\
 422 & 0 & \{\} \\
 \text{4221'} & 0 & \{2,2,2\} \\
\end{array}&&
\begin{array}{lll}
\mbox{MPG}&\mbox{Free}&\mbox{Tor}\\
\hline 
 \text{4'22'} & 0 & \{2\} \\
 \text{42'2'} & 0 & \{2\} \\
 \text{4mm} & 2 & \{\} \\
 \text{4mm1'} & 2 & \{\} \\
 \text{4'm'm} & 1 & \{\} \\
 \text{4m'm'} & 0 & \{\} \\
 \bar{4}\text{2m} & 1 & \{\} \\
 \bar{4}\text{2m1'} & 1 & \{2\} \\
 \bar{4}\text{'2'm} & 1 & \{\} \\
 \bar{4}\text{'2m'} & 0 & \{\} \\
 \bar{4}\text{2'm'} & 0 & \{2\} \\
 \text{4/mmm} & 3 & \{\} \\
 \text{4/mmm1'} & 3 & \{\} \\
 \text{4/m'mm} & 2 & \{\} \\
 \text{4'/mm'm} & 2 & \{\} \\
 \text{4'/m'm'm} & 1 & \{\} \\
 \text{4/mm'm'} & 1 & \{\} \\
 \text{4/m'm'm'} & 0 & \{\} \\
 3 & 0 & \{\} \\
 \text{31'} & 0 & \{2\} \\
 \bar{3} & 0 & \{2\} \\
 \bar{3}\text{1'} & 0 & \{4\} \\
 \bar{3}' & 0 & \{\} \\
 32 & 0 & \{\} \\
 \text{321'} & 0 & \{2,2\} \\
 \text{32'} & 0 & \{2\} \\
 \text{3m} & 1 & \{\} \\
 \text{3m1'} & 1 & \{\} \\
 \text{3m'} & 0 & \{\} \\
 \bar{3}\text{m} & 1 & \{\} \\
 \bar{3}\text{m1'} & 1 & \{2\} \\
 \bar{3}\text{'m} & 1 & \{\} \\
 \bar{3}\text{'m'} & 0 & \{\} \\
 \bar{3}\text{m'} & 0 & \{2\} \\
 6 & 0 & \{\} \\
 \text{61'} & 0 & \{2,2\} \\
 \text{6'} & 0 & \{2\} \\
 \bar{6} & 1 & \{\} \\
 \bar{6}\text{1'} & 1 & \{\} \\
 \bar{6}' & 0 & \{\} \\
 \text{6/m} & 1 & \{\} \\
\end{array}&&
\begin{array}{lll}
\mbox{MPG}&\mbox{Free}&\mbox{Tor}\\
\hline 
 \text{6/m1'} & 1 & \{2\} \\
 \text{6'/m} & 1 & \{\} \\
 \text{6/m'} & 0 & \{\} \\
 \text{6'/m'} & 0 & \{2\} \\
 622 & 0 & \{\} \\
 \text{6221'} & 0 & \{2,2,2\} \\
 \text{6'22'} & 0 & \{2\} \\
 \text{62'2'} & 0 & \{2\} \\
 \text{6mm} & 2 & \{\} \\
 \text{6mm1'} & 2 & \{\} \\
 \text{6'mm'} & 1 & \{\} \\
 \text{6m'm'} & 0 & \{\} \\
 \bar{6}\text{m2} & 2 & \{\} \\
 \bar{6}\text{m21'} & 2 & \{\} \\
 \bar{6}\text{'m'2} & 0 & \{\} \\
 \bar{6}\text{'m2'} & 1 & \{\} \\
 \bar{6}\text{m'2'} & 1 & \{\} \\
 \text{6/mmm} & 3 & \{\} \\
 \text{6/mmm1'} & 3 & \{\} \\
 \text{6/m'mm} & 2 & \{\} \\
 \text{6'/mmm'} & 2 & \{\} \\
 \text{6'/m'mm'} & 1 & \{\} \\
 \text{6/mm'm'} & 1 & \{\} \\
 \text{6/m'm'm'} & 0 & \{\} \\
 23 & 0 & \{\} \\
 \text{231'} & 0 & \{2\} \\
 \text{m}\bar{3} & 1 & \{\} \\
 \text{m}\bar{3}\text{1'} & 1 & \{\} \\
 \text{m'}\bar{3}' & 0 & \{\} \\
 432 & 0 & \{\} \\
 \text{4321'} & 0 & \{2,2\} \\
 \text{4'32'} & 0 & \{2\} \\
 \bar{4}\text{3m} & 1 & \{\} \\
 \bar{4}\text{3m1'} & 1 & \{\} \\
 \bar{4}\text{'3m'} & 0 & \{\} \\
 \text{m}\bar{3}\text{m} & 2 & \{\} \\
 \text{m}\bar{3}\text{m1'} & 2 & \{\} \\
 \text{m'}\bar{3}\text{'m} & 1 & \{\} \\
 \text{m}\bar{3}\text{m'} & 1 & \{\} \\
 \text{m'}\bar{3}\text{'m'} & 0 & \{\} \\
\ \\
\end{array}
\end{align*}
\end{table}


\renewcommand{\arraystretch}{1.0}
\begin{table}
\caption{EAZ classes of superconductors of spinless electrons with crystallographic MPG symmetry.}
\label{tab:EAZ_SC_spinless}
\fontsize{7pt}{9pt}\selectfont
\begin{align*}
\begin{array}{llll}
\mbox{MPG}&G_u&\mbox{Irrep}&\mbox{EAZ}\\
\hline 
 1 & 1 & \text{A} & \{\text{D}\} \\
 \text{11'} & 1 & \text{A} & \{\text{BDI}\} \\
 \bar{1} & \bar{1} & \text{A}_{\text{g}} & \{\text{BDI}\} \\
 \bar{1} & \bar{1} & \text{A}_{\text{u}} & \{\text{DIII}\} \\
 \bar{1}\text{1'} & \bar{1} & \text{A}_{\text{g}} & \{\text{AI}_C\} \\
 \bar{1}\text{1'} & \bar{1} & \text{A}_{\text{u}} & \{\text{AIII}_T\} \\
 \bar{1}' & 1 & \text{A} & \{\text{A}_C\} \\
\end{array}&&
\begin{array}{llll}
\mbox{MPG}&G_u&\mbox{Irrep}&\mbox{EAZ}\\
\hline 
 \text{m'mm} & \text{2mm} & \text{A}_2 & \{\text{A}_{T,C}\} \\
 \text{m'mm} & \text{2mm} & \text{B}_2 & \{\text{DIII}\} \\
 \text{m'mm} & \text{2mm} & \text{B}_1 & \{\text{DIII}\} \\
 \text{m'm'm} & \text{112/m} & \text{A}_{\text{g}} & \{\text{DIII}\} \\
 \text{m'm'm} & \text{112/m} & \text{A}_{\text{u}} & \{\text{DIII}\} \\
 \text{m'm'm} & \text{112/m} & \text{B}_{\text{g}} & \left\{\text{D}_T{}^2\right\} \\
 \text{m'm'm} & \text{112/m} & \text{B}_{\text{u}} & \left\{\text{DIII}^4\right\} \\
\end{array}&&
  \begin{array}{llll}
\mbox{MPG}&G_u&\mbox{Irrep}&\mbox{EAZ}\\
\hline 
 \text{4m'm'} & 4 & {}^1\text{E} & \left\{\text{D}^5\right\} \\
 \bar{4}\text{2m} & \bar{4}2\text{m}_{110} & \text{A}_1 & \{\text{CII}\} \\
 \bar{4}\text{2m} & \bar{4}2\text{m}_{110} & \text{B}_1 & \{\text{CI}\} \\
 \bar{4}\text{2m} & \bar{4}2\text{m}_{110} & \text{B}_2 & \{\text{DIII}\} \\
 \bar{4}\text{2m} & \bar{4}2\text{m}_{110} & \text{A}_2 & \{\text{BDI}\} \\
 \bar{4}\text{2m1'} & \bar{4}2\text{m}_{110} & \text{A}_1 & \{\text{AII}_C\} \\
 \bar{4}\text{2m1'} & \bar{4}2\text{m}_{110} & \text{B}_1 & \{\text{AIII}_T\} \\
\end{array}
\end{align*}
\end{table}

\begin{table*}
\caption*{(Continued from the previous page)}
\fontsize{7pt}{9pt}\selectfont
\begin{align*}
\begin{array}{llll}
\mbox{MPG}&G_u&\mbox{Irrep}&\mbox{EAZ}\\
\hline 
 32 & 32_{100}1 & \text{A}_1 & \{\text{C},\text{A}_C\} \\
 32 & 32_{100}1 & \text{A}_2 & \left\{\text{D}^3\right\} \\
 \text{321'} & 32_{100}1 & \text{A}_1 & \{\text{AIII}_T,\text{CII}\} \\
 \text{321'} & 32_{100}1 & \text{A}_2 & \{\text{D}_T,\text{DIII}\} \\
 \text{32'} & 3 & \text{A} & \{\text{DIII},\text{AII}_C\} \\
 \text{32'} & 3 & {}^2\text{E} & \{\text{DIII},\text{AII}_C\} \\
 \text{32'} & 3 & {}^1\text{E} & \{\text{DIII},\text{AII}_C\} \\
\end{array}&&
\begin{array}{llll}
\mbox{MPG}&G_u&\mbox{Irrep}&\mbox{EAZ}\\
\hline 
 \text{6'/m'} & \bar{3} & \text{A}_{\text{g}} & \{\text{A}_{T,C},\text{D}_T\} \\
 \text{6'/m'} & \bar{3} & \text{A}_{\text{u}} & \left\{\text{AIII}_T{}^2,\text{DIII}^2\right\} \\
 622 & 622 & \text{A}_1 & \left\{\text{C}^3\right\} \\
 622 & 622 & \text{B}_2 & \{\text{D},\text{A}_C\} \\
 622 & 622 & \text{B}_1 & \{\text{D},\text{A}_C\} \\
 622 & 622 & \text{A}_2 & \left\{\text{D}^3\right\} \\
 \text{6221'} & 622 & \text{A}_1 & \left\{\text{CII}^3\right\} \\
\end{array}&&
\begin{array}{llll}
\mbox{MPG}&G_u&\mbox{Irrep}&\mbox{EAZ}\\
\hline 
 \text{6/m'mm} & \text{6mm} & \text{B}_1 & \{\text{DIII},\text{AII}_C\} \\
 \text{6/m'mm} & \text{6mm} & \text{B}_2 & \{\text{DIII},\text{AII}_C\} \\
 \text{6'/mmm'} & \bar{6}2_{210}\text{m}_{100} & \text{A'}_1 & \left\{\text{AII}_C{}^3\right\} \\
 \text{6'/mmm'} & \bar{6}2_{210}\text{m}_{100} & \text{A''}_1 & \{\text{A}_{T,C},\text{AIII}_T\} \\
 \text{6'/mmm'} & \bar{6}2_{210}\text{m}_{100} & \text{A''}_2 & \left\{\text{DIII}^3\right\} \\
 \text{6'/mmm'} & \bar{6}2_{210}\text{m}_{100} & \text{A'}_2 & \{\text{D}_T,\text{DIII}\} \\
 \text{6'/m'mm'} & \bar{3}\text{m}_{100} & \text{A}_{\text{1g}} & \{\text{DIII},\text{AII}_C\} \\
\end{array}
\end{align*}
\end{table*}

\begin{table}
\caption{EAZ classes of superconductors of spinful electrons with crystallographic MPG symmetry.}
\label{tab:EAZ_SC_spinful}
\fontsize{7pt}{9pt}\selectfont
\begin{align*}
\begin{array}{llll}
\mbox{MPG}&G_u&\mbox{Irrep}&\mbox{EAZ}\\
\hline 
 1 & 1 & \text{A} & \{\text{D}\} \\
 \text{11'} & 1 & \text{A} & \{\text{DIII}\} \\
 \bar{1} & \bar{1} & \text{A}_{\text{g}} & \{\text{BDI}\} \\
 \bar{1} & \bar{1} & \text{A}_{\text{u}} & \{\text{DIII}\} \\
 \bar{1}\text{1'} & \bar{1} & \text{A}_{\text{g}} & \{\text{D}_T\} \\
 \bar{1}\text{1'} & \bar{1} & \text{A}_{\text{u}} & \left\{\text{DIII}^2\right\} \\
 \bar{1}' & 1 & \text{A} & \left\{\text{D}^2\right\} \\
\end{array}&&
\begin{array}{llll}
\mbox{MPG}&G_u&\mbox{Irrep}&\mbox{EAZ}\\
\hline 
 \text{m'mm} & \text{2mm} & \text{A}_2 & \left\{\text{DIII}^4\right\} \\
 \text{m'mm} & \text{2mm} & \text{B}_2 & \{\text{DIII}\} \\
 \text{m'mm} & \text{2mm} & \text{B}_1 & \{\text{DIII}\} \\
 \text{m'm'm} & \text{112/m} & \text{A}_{\text{g}} & \left\{\text{D}_T{}^2\right\} \\
 \text{m'm'm} & \text{112/m} & \text{A}_{\text{u}} & \left\{\text{DIII}^4\right\} \\
 \text{m'm'm} & \text{112/m} & \text{B}_{\text{g}} & \{\text{DIII}\} \\
 \text{m'm'm} & \text{112/m} & \text{B}_{\text{u}} & \{\text{DIII}\} \\
\end{array}&&
 \begin{array}{llll}
\mbox{MPG}&G_u&\mbox{Irrep}&\mbox{EAZ}\\
\hline 
 \text{4m'm'} & 4 & {}^1\text{E} & \left\{\text{D}^2\right\} \\
 \bar{4}\text{2m} & \bar{4}2\text{m}_{110} & \text{A}_1 & \left\{\text{BDI}^2,\text{D}_T\right\} \\
 \bar{4}\text{2m} & \bar{4}2\text{m}_{110} & \text{B}_1 & \left\{\text{D}_T,\text{DIII}^2\right\} \\
 \bar{4}\text{2m} & \bar{4}2\text{m}_{110} & \text{B}_2 & \{\text{AIII}_T,\text{AI}_C\} \\
 \bar{4}\text{2m} & \bar{4}2\text{m}_{110} & \text{A}_2 & \{\text{AIII}_T,\text{AII}_C\} \\
 \bar{4}\text{2m1'} & \bar{4}2\text{m}_{110} & \text{A}_1 & \left\{\text{D}_T{}^2,\text{DIII}\right\} \\
 \bar{4}\text{2m1'} & \bar{4}2\text{m}_{110} & \text{B}_1 & \left\{\text{DIII}^5\right\} \\
\end{array}
\end{align*}
\end{table}

\begin{table*}
\caption*{(Continued from the previous page)}
\fontsize{7pt}{9pt}\selectfont
\begin{align*}
\begin{array}{llll}
\mbox{MPG}&G_u&\mbox{Irrep}&\mbox{EAZ}\\
\hline 
 32 & 32_{100}1 & \text{A}_1 & \left\{\text{D}^3\right\} \\
 32 & 32_{100}1 & \text{A}_2 & \{\text{C},\text{A}_C\} \\
 \text{321'} & 32_{100}1 & \text{A}_1 & \left\{\text{DIII}^3\right\} \\
 \text{321'} & 32_{100}1 & \text{A}_2 & \{\text{AII}_C,\text{CII}\} \\
 \text{32'} & 3 & \text{A} & \{\text{DIII},\text{AII}_C\} \\
 \text{32'} & 3 & {}^2\text{E} & \{\text{DIII},\text{AII}_C\} \\
 \text{32'} & 3 & {}^1\text{E} & \{\text{DIII},\text{AII}_C\} \\
\end{array}&&
\begin{array}{llll}
\mbox{MPG}&G_u&\mbox{Irrep}&\mbox{EAZ}\\
\hline 
 \text{6'/m'} & \bar{3} & \text{A}_{\text{g}} & \{\text{A}_{T,C},\text{D}_T\} \\
 \text{6'/m'} & \bar{3} & \text{A}_{\text{u}} & \left\{\text{AIII}_T{}^2,\text{DIII}^2\right\} \\
 622 & 622 & \text{A}_1 & \left\{\text{D}^6\right\} \\
 622 & 622 & \text{B}_2 & \left\{\text{A}_C{}^3\right\} \\
 622 & 622 & \text{B}_1 & \left\{\text{A}_C{}^3\right\} \\
 622 & 622 & \text{A}_2 & \left\{\text{C}^2,\text{A}_C{}^2\right\} \\
 \text{6221'} & 622 & \text{A}_1 & \left\{\text{DIII}^6\right\} \\
\end{array}&&
\begin{array}{llll}
\mbox{MPG}&G_u&\mbox{Irrep}&\mbox{EAZ}\\
\hline 
 \text{6/m'mm} & \text{6mm} & \text{B}_1 & \{\text{DIII},\text{AII}_C\} \\
 \text{6/m'mm} & \text{6mm} & \text{B}_2 & \{\text{DIII},\text{AII}_C\} \\
 \text{6'/mmm'} & \bar{6}2_{210}\text{m}_{100} & \text{A'}_1 & \left\{\text{D}_T{}^3\right\} \\
 \text{6'/mmm'} & \bar{6}2_{210}\text{m}_{100} & \text{A''}_1 & \left\{\text{DIII}^6\right\} \\
 \text{6'/mmm'} & \bar{6}2_{210}\text{m}_{100} & \text{A''}_2 & \{\text{AIII}_T,\text{DIII}\} \\
 \text{6'/mmm'} & \bar{6}2_{210}\text{m}_{100} & \text{A'}_2 & \{\text{DIII},\text{AII}_C\} \\
 \text{6'/m'mm'} & \bar{3}\text{m}_{100} & \text{A}_{\text{1g}} & \left\{\text{D}_T{}^3\right\} \\
\end{array}
\end{align*}
\end{table*}

\begin{table}
\caption{The classification of surface states of $3d$ TSCs of spinless electrons with crystallographic MPG symmetry.}
\label{tab:surf_states_sc_spinless}
\fontsize{7pt}{9pt}\selectfont
\begin{align*}
\begin{array}{lllll}
\mbox{MPG}&G_u&\mbox{Irrep}&\mbox{Free}&\mbox{Tor}\\
\hline 
 1 & 1 & \text{A} & 0 & \{\} \\
 \text{11'} & 1 & \text{A} & 0 & \{\} \\
 \bar{1} & \bar{1} & \text{A}_{\text{g}} & 0 & \{\} \\
 \bar{1} & \bar{1} & \text{A}_{\text{u}} & 0 & \{4\} \\
 \bar{1}\text{1'} & \bar{1} & \text{A}_{\text{g}} & 0 & \{\} \\
 \bar{1}\text{1'} & \bar{1} & \text{A}_{\text{u}} & 0 & \{2\} \\
 \bar{1}' & 1 & \text{A} & 0 & \{\} \\
\end{array}&&
\begin{array}{lllll}
\mbox{MPG}&G_u&\mbox{Irrep}&\mbox{Free}&\mbox{Tor}\\
\hline 
 \text{m'mm} & \text{2mm} & \text{A}_2 & 0 & \{\} \\
 \text{m'mm} & \text{2mm} & \text{B}_2 & 1 & \{\} \\
 \text{m'mm} & \text{2mm} & \text{B}_1 & 1 & \{\} \\
 \text{m'm'm} & \text{112/m} & \text{A}_{\text{g}} & 1 & \{\} \\
 \text{m'm'm} & \text{112/m} & \text{A}_{\text{u}} & 0 & \{2\} \\
 \text{m'm'm} & \text{112/m} & \text{B}_{\text{g}} & 0 & \{\} \\
 \text{m'm'm} & \text{112/m} & \text{B}_{\text{u}} & 3 & \{\} \\
\end{array}&&
 \begin{array}{lllll}
\mbox{MPG}&G_u&\mbox{Irrep}&\mbox{Free}&\mbox{Tor}\\
\hline 
 \text{4m'm'} & 4 & {}^1\text{E} & 0 & \{\} \\
 \bar{4}\text{2m} & \bar{4}2\text{m}_{110} & \text{A}_1 & 0 & \{\} \\
 \bar{4}\text{2m} & \bar{4}2\text{m}_{110} & \text{B}_1 & 0 & \{\} \\
 \bar{4}\text{2m} & \bar{4}2\text{m}_{110} & \text{B}_2 & 0 & \{2\} \\
 \bar{4}\text{2m} & \bar{4}2\text{m}_{110} & \text{A}_2 & 0 & \{\} \\
 \bar{4}\text{2m1'} & \bar{4}2\text{m}_{110} & \text{A}_1 & 0 & \{\} \\
 \bar{4}\text{2m1'} & \bar{4}2\text{m}_{110} & \text{B}_1 & 0 & \{\} \\
\end{array}
\end{align*}
\end{table}

\begin{table*}
\caption*{(Continued from the previous page)}
\fontsize{7pt}{9pt}\selectfont
\begin{align*}
\begin{array}{lllll}
\mbox{MPG}&G_u&\mbox{Irrep}&\mbox{Free}&\mbox{Tor}\\
\hline 
 32 & 32_{100}1 & \text{A}_1 & 0 & \{\} \\
 32 & 32_{100}1 & \text{A}_2 & 0 & \{\} \\
 \text{321'} & 32_{100}1 & \text{A}_1 & 1 & \{\} \\
 \text{321'} & 32_{100}1 & \text{A}_2 & 1 & \{\} \\
 \text{32'} & 3 & \text{A} & 1 & \{\} \\
 \text{32'} & 3 & {}^2\text{E} & 1 & \{\} \\
 \text{32'} & 3 & {}^1\text{E} & 1 & \{\} \\
\end{array}&&
\begin{array}{lllll}
\mbox{MPG}&G_u&\mbox{Irrep}&\mbox{Free}&\mbox{Tor}\\
\hline 
 \text{6'/m'} & \bar{3} & \text{A}_{\text{g}} & 0 & \{\} \\
 \text{6'/m'} & \bar{3} & \text{A}_{\text{u}} & 2 & \{2\} \\
 622 & 622 & \text{A}_1 & 0 & \{\} \\
 622 & 622 & \text{B}_2 & 0 & \{\} \\
 622 & 622 & \text{B}_1 & 0 & \{\} \\
 622 & 622 & \text{A}_2 & 0 & \{\} \\
 \text{6221'} & 622 & \text{A}_1 & 0 & \{\} \\
\end{array}&&
\begin{array}{lllll}
\mbox{MPG}&G_u&\mbox{Irrep}&\mbox{Free}&\mbox{Tor}\\
\hline 
 \text{6/m'mm} & \text{6mm} & \text{B}_1 & 1 & \{\} \\
 \text{6/m'mm} & \text{6mm} & \text{B}_2 & 1 & \{\} \\
 \text{6'/mmm'} & \bar{6}2_{210}\text{m}_{100} & \text{A'}_1 & 0 & \{2\} \\
 \text{6'/mmm'} & \bar{6}2_{210}\text{m}_{100} & \text{A''}_1 & 0 & \{\} \\
 \text{6'/mmm'} & \bar{6}2_{210}\text{m}_{100} & \text{A''}_2 & 2 & \{\} \\
 \text{6'/mmm'} & \bar{6}2_{210}\text{m}_{100} & \text{A'}_2 & 1 & \{\} \\
 \text{6'/m'mm'} & \bar{3}\text{m}_{100} & \text{A}_{\text{1g}} & 1 & \{\} \\
\end{array}
\end{align*}
\end{table*}

\begin{table}
\caption{The classification of surface states of $3d$ TSCs of spinful electrons with crystallographic MPG symmetry. 
The bold red characters denoted by $\textcolor{red}{\bf 1}$ represent the 1st-order TSC.}
\label{tab:surf_states_sc_spinful}
\fontsize{7pt}{9pt}\selectfont
\begin{align*}
\begin{array}{lllll}
\mbox{MPG}&G_u&\mbox{Irrep}&\mbox{Free}&\mbox{Tor}\\
\hline 
 1 & 1 & \text{A} & 0 & \{\} \\
 \text{11'} & 1 & \text{A} & \text{\textcolor{red}{\bf 1}} & \{\} \\
 \bar{1} & \bar{1} & \text{A}_{\text{g}} & 0 & \{\} \\
 \bar{1} & \bar{1} & \text{A}_{\text{u}} & 0 & \{4\} \\
 \bar{1}\text{1'} & \bar{1} & \text{A}_{\text{g}} & 0 & \{\} \\
 \bar{1}\text{1'} & \bar{1} & \text{A}_{\text{u}} & \text{\textcolor{red}{\bf 1}} & \{4\} \\
 \bar{1}' & 1 & \text{A} & 0 & \{\} \\
\end{array}&&
\begin{array}{lllll}
\mbox{MPG}&G_u&\mbox{Irrep}&\mbox{Free}&\mbox{Tor}\\
\hline 
 \text{m'mm} & \text{2mm} & \text{A}_2 & 4 & \{\} \\
 \text{m'mm} & \text{2mm} & \text{B}_2 & 1 & \{\} \\
 \text{m'mm} & \text{2mm} & \text{B}_1 & 1 & \{\} \\
 \text{m'm'm} & \text{112/m} & \text{A}_{\text{g}} & 0 & \{\} \\
 \text{m'm'm} & \text{112/m} & \text{A}_{\text{u}} & 3 & \{\} \\
 \text{m'm'm} & \text{112/m} & \text{B}_{\text{g}} & 1 & \{\} \\
 \text{m'm'm} & \text{112/m} & \text{B}_{\text{u}} & 0 & \{2\} \\
\end{array}&&
 \begin{array}{lllll}
\mbox{MPG}&G_u&\mbox{Irrep}&\mbox{Free}&\mbox{Tor}\\
\hline 
 \text{4m'm'} & 4 & {}^1\text{E} & 0 & \{\} \\
 \bar{4}\text{2m} & \bar{4}2\text{m}_{110} & \text{A}_1 & 0 & \{\} \\
 \bar{4}\text{2m} & \bar{4}2\text{m}_{110} & \text{B}_1 & 1 & \{2\} \\
 \bar{4}\text{2m} & \bar{4}2\text{m}_{110} & \text{B}_2 & 0 & \{2\} \\
 \bar{4}\text{2m} & \bar{4}2\text{m}_{110} & \text{A}_2 & 1 & \{2\} \\
 \bar{4}\text{2m1'} & \bar{4}2\text{m}_{110} & \text{A}_1 & 1 & \{\} \\
 \bar{4}\text{2m1'} & \bar{4}2\text{m}_{110} & \text{B}_1 & \text{\textcolor{red}{\bf 1}+3} & \{\} \\
\end{array}
\end{align*}
\end{table}

\begin{table*}
\caption*{(Continued from the previous page)}
\fontsize{7pt}{9pt}\selectfont
\begin{align*}
\begin{array}{lllll}
\mbox{MPG}&G_u&\mbox{Irrep}&\mbox{Free}&\mbox{Tor}\\
\hline 
 32 & 32_{100}1 & \text{A}_1 & 0 & \{\} \\
 32 & 32_{100}1 & \text{A}_2 & 0 & \{\} \\
 \text{321'} & 32_{100}1 & \text{A}_1 & \text{\textcolor{red}{\bf 1}+2} & \{\} \\
 \text{321'} & 32_{100}1 & \text{A}_2 & 0 & \{2\} \\
 \text{32'} & 3 & \text{A} & 1 & \{\} \\
 \text{32'} & 3 & {}^2\text{E} & 1 & \{\} \\
 \text{32'} & 3 & {}^1\text{E} & 1 & \{\} \\
\end{array}&&
\begin{array}{lllll}
\mbox{MPG}&G_u&\mbox{Irrep}&\mbox{Free}&\mbox{Tor}\\
\hline 
 \text{6'/m'} & \bar{3} & \text{A}_{\text{g}} & 0 & \{\} \\
 \text{6'/m'} & \bar{3} & \text{A}_{\text{u}} & 2 & \{2\} \\
 622 & 622 & \text{A}_1 & 0 & \{\} \\
 622 & 622 & \text{B}_2 & 0 & \{\} \\
 622 & 622 & \text{B}_1 & 0 & \{\} \\
 622 & 622 & \text{A}_2 & 0 & \{\} \\
 \text{6221'} & 622 & \text{A}_1 & \text{\textcolor{red}{\bf 1}+5} & \{\} \\
\end{array}&&
\begin{array}{lllll}
\mbox{MPG}&G_u&\mbox{Irrep}&\mbox{Free}&\mbox{Tor}\\
\hline 
 \text{6/m'mm} & \text{6mm} & \text{B}_1 & 1 & \{\} \\
 \text{6/m'mm} & \text{6mm} & \text{B}_2 & 1 & \{\} \\
 \text{6'/mmm'} & \bar{6}2_{210}\text{m}_{100} & \text{A'}_1 & 0 & \{\} \\
 \text{6'/mmm'} & \bar{6}2_{210}\text{m}_{100} & \text{A''}_1 & 5 & \{\} \\
 \text{6'/mmm'} & \bar{6}2_{210}\text{m}_{100} & \text{A''}_2 & 1 & \{\} \\
 \text{6'/mmm'} & \bar{6}2_{210}\text{m}_{100} & \text{A'}_2 & 1 & \{\} \\
 \text{6'/m'mm'} & \bar{3}\text{m}_{100} & \text{A}_{\text{1g}} & 0 & \{\} \\
\end{array}
\end{align*}
\end{table*}

\begin{table}
\caption{Character tables of 1-dimensional representations of the crystallographic point groups.
\label{tab:1Drep}
}
\fontsize{7pt}{9pt}\selectfont
\begin{align*}
\begin{array}{c|c}
 1 & 1 \\
 \hline
 \text{A} & 1 \\
\end{array}
&&
\begin{array}{c|cc}
 211 & 1 & 2_{100} \\
 \hline
 \text{A} & 1 & 1 \\
 \text{B} & 1 & -1 \\
\end{array}
&&
\begin{array}{c|cc}
 121 & 1 & 2_{010} \\
 \hline
 \text{A} & 1 & 1 \\
 \text{B} & 1 & -1 \\
\end{array}
&&
\begin{array}{c|cc}
 112 & 1 & 2_{001} \\
 \hline
 \text{A} & 1 & 1 \\
 \text{B} & 1 & -1 \\
\end{array}
&&
\begin{array}{c|cc}
 \bar{1} & 1 & \bar{1} \\\hline
 \text{A}_{\text{g}} & 1 & 1 \\
 \text{A}_{\text{u}} & 1 & -1 \\
\end{array}
&&
\begin{array}{c|cc}
 \text{m11} & 1 & \text{m}_{100} \\\hline
 \text{A'} & 1 & 1 \\
 \text{A''} & 1 & -1 \\
\end{array}
&&
\begin{array}{c|cc}
 \text{1m1} & 1 & \text{m}_{010} \\\hline
 \text{A'} & 1 & 1 \\
 \text{A''} & 1 & -1 \\
\end{array}
\end{align*}
\begin{align*}
\begin{array}{c|cc}
 \text{11m} & 1 & \text{m}_{001} \\\hline
 \text{A'} & 1 & 1 \\
 \text{A''} & 1 & -1 \\
\end{array}
&&
\begin{array}{c|ccc}
 3 & 1 & 3^+_{001} & 3^-_{001} \\\hline
 \text{A} & 1 & 1 & 1 \\
 {}^2\text{E} & 1 & e^{\frac{2 \pi i}{3}} & e^{-\frac{2 \pi i}{3}} \\
 {}^1\text{E} & 1 & e^{-\frac{2 \pi i}{3}} & e^{\frac{2 \pi i}{3}} \\
\end{array}
&&
\begin{array}{c|cccc}
 222 & 1 & 2_{100} & 2_{010} & 2_{001} \\\hline
 \text{A}_1 & 1 & 1 & 1 & 1 \\
 \text{B}_3 & 1 & 1 & -1 & -1 \\
 \text{B}_2 & 1 & -1 & 1 & -1 \\
 \text{B}_1 & 1 & -1 & -1 & 1 \\
\end{array}
&&
\begin{array}{c|cccc}
 \text{2/m11} & 1 & 2_{100} & \bar{1} & \text{m}_{100} \\\hline
 \text{A}_{\text{g}} & 1 & 1 & 1 & 1 \\
 \text{A}_{\text{u}} & 1 & 1 & -1 & -1 \\
 \text{B}_{\text{g}} & 1 & -1 & 1 & -1 \\
 \text{B}_{\text{u}} & 1 & -1 & -1 & 1 \\
\end{array}
\end{align*}
\begin{align*}
\begin{array}{c|cccc}
 \text{2mm} & 1 & 2_{100} & \text{m}_{010} & \text{m}_{001} \\\hline
 \text{A}_1 & 1 & 1 & 1 & 1 \\
 \text{A}_2 & 1 & 1 & -1 & -1 \\
 \text{B}_2 & 1 & -1 & 1 & -1 \\
 \text{B}_1 & 1 & -1 & -1 & 1 \\
\end{array}
&&
\begin{array}{c|cccc}
 \text{12/m1} & 1 & 2_{010} & \bar{1} & \text{m}_{010} \\\hline
 \text{A}_{\text{g}} & 1 & 1 & 1 & 1 \\
 \text{A}_{\text{u}} & 1 & 1 & -1 & -1 \\
 \text{B}_{\text{g}} & 1 & -1 & 1 & -1 \\
 \text{B}_{\text{u}} & 1 & -1 & -1 & 1 \\
\end{array}
&&
\begin{array}{c|cccc}
 \text{m2m} & 1 & 2_{010} & \text{m}_{100} & \text{m}_{001} \\\hline
 \text{A}_1 & 1 & 1 & 1 & 1 \\
 \text{A}_2 & 1 & 1 & -1 & -1 \\
 \text{B}_1 & 1 & -1 & 1 & -1 \\
 \text{B}_2 & 1 & -1 & -1 & 1 \\
\end{array}
\end{align*}
\begin{align*}
\begin{array}{c|cccc}
 2_{110}2_{1\bar{1}0}2 & 1 & 2_{001} & 2_{1\bar{1}0} & 2_{110} \\\hline
 \text{A}_1 & 1 & 1 & 1 & 1 \\
 \text{B}_1 & 1 & 1 & -1 & -1 \\
\end{array}
&&
\begin{array}{c|cccc}
 4 & 1 & 2_{001} & 4^+_{001} & 4^-_{001} \\\hline
 \text{A} & 1 & 1 & 1 & 1 \\
 \text{B} & 1 & 1 & -1 & -1 \\
 {}^2\text{E} & 1 & -1 & i & -i \\
 {}^1\text{E} & 1 & -1 & -i & i \\
\end{array}
&&
\begin{array}{c|cccc}
 \text{112/m} & 1 & 2_{001} & \bar{1} & \text{m}_{001} \\\hline
 \text{A}_{\text{g}} & 1 & 1 & 1 & 1 \\
 \text{A}_{\text{u}} & 1 & 1 & -1 & -1 \\
 \text{B}_{\text{g}} & 1 & -1 & 1 & -1 \\
 \text{B}_{\text{u}} & 1 & -1 & -1 & 1 \\
\end{array}
\end{align*}
\begin{align*}
\begin{array}{c|cccc}
 \text{mm2} & 1 & 2_{001} & \text{m}_{100} & \text{m}_{010} \\\hline
 \text{A}_1 & 1 & 1 & 1 & 1 \\
 \text{A}_2 & 1 & 1 & -1 & -1 \\
 \text{B}_2 & 1 & -1 & 1 & -1 \\
 \text{B}_1 & 1 & -1 & -1 & 1 \\
\end{array}
&&
\begin{array}{c|cccc}
 \text{m}_{110}\text{m}_{1-10}2 & 1 & 2_{001} & \text{m}_{1\bar{1}0} & \text{m}_{110} \\\hline
 \text{A}_1 & 1 & 1 & 1 & 1 \\
 \text{A}_2 & 1 & 1 & -1 & -1 \\
\end{array}
&&
\begin{array}{c|cccc}
 \bar{4} & 1 & 2_{001} & \bar{4}^+_{001} & \bar{4}^-_{001} \\\hline
 \text{A} & 1 & 1 & 1 & 1 \\
 \text{B} & 1 & 1 & -1 & -1 \\
 {}^2\text{E} & 1 & -1 & i & -i \\
 {}^1\text{E} & 1 & -1 & -i & i \\
\end{array}
\end{align*}
\begin{align*}
\begin{array}{c|cccccc}
 32_{100}1 & 1 & 2_{100} & 3^+_{001} & 3^-_{001} & 2_{110} & 2_{010} \\\hline
 \text{A}_1 & 1 & 1 & 1 & 1 & 1 & 1 \\
 \text{A}_2 & 1 & -1 & 1 & 1 & -1 & -1 \\
\end{array}
&&
\begin{array}{c|cccccc}
 312_{120} & 1 & 2_{120} & 3^+_{001} & 3^-_{001} & 2_{210} & 2_{1\bar{1}0} \\\hline
 \text{A}_1 & 1 & 1 & 1 & 1 & 1 & 1 \\
 \text{A}_2 & 1 & -1 & 1 & 1 & -1 & -1 \\
\end{array}
\end{align*}
\begin{align*}
\begin{array}{c|cccccc}
 6 & 1 & 2_{001} & 6^+_{001} & 3^+_{001} & 3^-_{001} & 6^-_{001} \\\hline
 \text{A} & 1 & 1 & 1 & 1 & 1 & 1 \\
 {}^1\text{E}_2 & 1 & 1 & e^{\frac{2 \pi i}{3}} & e^{-\frac{2 \pi i}{3}} & e^{\frac{2 \pi i}{3}} & e^{-\frac{2 \pi i}{3}} \\
 {}^2\text{E}_2 & 1 & 1 & e^{-\frac{2 \pi i}{3}} & e^{\frac{2 \pi i}{3}} & e^{-\frac{2 \pi i}{3}} & e^{\frac{2 \pi i}{3}} \\
 {}^2\text{E}_1 & 1 & -1 & e^{\frac{\pi i}{3}} & e^{\frac{2 \pi i}{3}} & e^{-\frac{2 \pi i}{3}} & e^{-\frac{\pi i}{3}} \\
 {}^1\text{E}_1 & 1 & -1 & e^{-\frac{\pi i}{3}} & e^{-\frac{2 \pi i}{3}} & e^{\frac{2 \pi i}{3}} & e^{\frac{\pi i}{3}} \\
 \text{B} & 1 & -1 & -1 & 1 & 1 & -1 \\
\end{array}
&&
\begin{array}{c|cccccc}
 \bar{3} & 1 & \bar{1} & 3^+_{001} & 3^-_{001} & \bar{3}^+_{001} & \bar{3}^-_{001} \\\hline
 \text{A}_{\text{g}} & 1 & 1 & 1 & 1 & 1 & 1 \\
 {}^2\text{E}_{\text{g}} & 1 & 1 & e^{\frac{2 \pi i}{3}} & e^{-\frac{2 \pi i}{3}} & e^{\frac{2 \pi i}{3}} & e^{-\frac{2 \pi i}{3}} \\
 {}^1\text{E}_{\text{g}} & 1 & 1 & e^{-\frac{2 \pi i}{3}} & e^{\frac{2 \pi i}{3}} & e^{-\frac{2 \pi i}{3}} & e^{\frac{2 \pi i}{3}} \\
 \text{A}_{\text{u}} & 1 & -1 & 1 & 1 & -1 & -1 \\
 {}^2\text{E}_{\text{u}} & 1 & -1 & e^{\frac{2 \pi i}{3}} & e^{-\frac{2 \pi i}{3}} & e^{-\frac{\pi i}{3}} & e^{\frac{\pi i}{3}} \\
 {}^1\text{E}_{\text{u}} & 1 & -1 & e^{-\frac{2 \pi i}{3}} & e^{\frac{2 \pi i}{3}} & e^{\frac{\pi i}{3}} & e^{-\frac{\pi i}{3}} \\
\end{array}
\end{align*}
\begin{align*}
\begin{array}{c|cccccc}
 31\text{m}_{100} & 1 & \text{m}_{100} & 3^+_{001} & 3^-_{001} & \text{m}_{110} & \text{m}_{010} \\\hline
 \text{A}_1 & 1 & 1 & 1 & 1 & 1 & 1 \\
 \text{A}_2 & 1 & -1 & 1 & 1 & -1 & -1 \\
\end{array}
&&
\begin{array}{c|cccccc}
 3\text{m}_{120}1 & 1 & \text{m}_{120} & 3^+_{001} & 3^-_{001} & \text{m}_{210} & \text{m}_{1\bar{1}0} \\\hline
 \text{A}_1 & 1 & 1 & 1 & 1 & 1 & 1 \\
 \text{A}_2 & 1 & -1 & 1 & 1 & -1 & -1 \\
\end{array}
\end{align*}
\begin{align*}
\begin{array}{c|cccccc}
 \bar{6} & 1 & \text{m}_{001} & 3^+_{001} & 3^-_{001} & \bar{6}^+_{001} & \bar{6}^-_{001} \\\hline
 \text{A'} & 1 & 1 & 1 & 1 & 1 & 1 \\
 {}^2\text{E'} & 1 & 1 & e^{\frac{2 \pi i}{3}} & e^{-\frac{2 \pi i}{3}} & e^{-\frac{2 \pi i}{3}} & e^{\frac{2 \pi i}{3}} \\
 {}^1\text{E'} & 1 & 1 & e^{-\frac{2 \pi i}{3}} & e^{\frac{2 \pi i}{3}} & e^{\frac{2 \pi i}{3}} & e^{-\frac{2 \pi i}{3}} \\
 \text{A''} & 1 & -1 & 1 & 1 & -1 & -1 \\
 {}^2\text{E''} & 1 & -1 & e^{\frac{2 \pi i}{3}} & e^{-\frac{2 \pi i}{3}} & e^{\frac{\pi i}{3}} & e^{-\frac{\pi i}{3}} \\
 {}^1\text{E''} & 1 & -1 & e^{-\frac{2 \pi i}{3}} & e^{\frac{2 \pi i}{3}} & e^{-\frac{\pi i}{3}} & e^{\frac{\pi i}{3}} \\
\end{array}
&&
\begin{array}{c|cccccccc}
 422 & 1 & 2_{100} & 2_{010} & 2_{001} & 2_{1\bar{1}0} & 4^+_{001} & 4^-_{001} & 2_{110} \\\hline
 \text{A}_1 & 1 & 1 & 1 & 1 & 1 & 1 & 1 & 1 \\
 \text{B}_1 & 1 & 1 & 1 & 1 & -1 & -1 & -1 & -1 \\
 \text{B}_2 & 1 & -1 & -1 & 1 & 1 & -1 & -1 & 1 \\
 \text{A}_2 & 1 & -1 & -1 & 1 & -1 & 1 & 1 & -1 \\
\end{array}
\end{align*}
\end{table}

\begin{table*}
\caption*{(Continued from the previous page)}
\fontsize{7pt}{9pt}\selectfont
\begin{align*}
\begin{array}{c|cccccccc}
 \text{mmm} & 1 & 2_{100} & 2_{010} & 2_{001} & \bar{1} & \text{m}_{100} & \text{m}_{010} & \text{m}_{001} \\\hline
 \text{A}_{\text{g}} & 1 & 1 & 1 & 1 & 1 & 1 & 1 & 1 \\
 \text{A}_{\text{u}} & 1 & 1 & 1 & 1 & -1 & -1 & -1 & -1 \\
 \text{B}_{\text{3g}} & 1 & 1 & -1 & -1 & 1 & 1 & -1 & -1 \\
 \text{B}_{\text{3u}} & 1 & 1 & -1 & -1 & -1 & -1 & 1 & 1 \\
 \text{B}_{\text{2g}} & 1 & -1 & 1 & -1 & 1 & -1 & 1 & -1 \\
 \text{B}_{\text{2u}} & 1 & -1 & 1 & -1 & -1 & 1 & -1 & 1 \\
 \text{B}_{\text{1g}} & 1 & -1 & -1 & 1 & 1 & -1 & -1 & 1 \\
 \text{B}_{\text{1u}} & 1 & -1 & -1 & 1 & -1 & 1 & 1 & -1 \\
\end{array}
&&
\begin{array}{c|cccccccc}
 \bar{4}2m_{110} & 1 & 2_{100} & 2_{010} & 2_{001} & \text{m}_{1\bar{1}0} & \bar{4}^+_{001} & \bar{4}^-_{001} & \text{m}_{110} \\\hline
 \text{A}_1 & 1 & 1 & 1 & 1 & 1 & 1 & 1 & 1 \\
 \text{B}_1 & 1 & 1 & 1 & 1 & -1 & -1 & -1 & -1 \\
 \text{B}_2 & 1 & -1 & -1 & 1 & 1 & -1 & -1 & 1 \\
 \text{A}_2 & 1 & -1 & -1 & 1 & -1 & 1 & 1 & -1 \\
\end{array}
\end{align*}
\begin{align*}
\begin{array}{c|cccccccc}
 \text{m}_{110}\text{m}_{1-10}\text{m} & 1 & 2_{001} & 2_{1\bar{1}0} & 2_{110} & \bar{1} & \text{m}_{001} & \text{m}_{1\bar{1}0} & \text{m}_{110} \\\hline
 \text{A}_{\text{g}} & 1 & 1 & 1 & 1 & 1 & 1 & 1 & 1 \\
 \text{A}_{\text{u}} & 1 & 1 & 1 & 1 & -1 & -1 & -1 & -1 \\
 \text{B}_{\text{1g}} & 1 & 1 & -1 & -1 & 1 & 1 & -1 & -1 \\
 \text{B}_{\text{1u}} & 1 & 1 & -1 & -1 & -1 & -1 & 1 & 1 \\
\end{array}
&&
\begin{array}{c|cccccccc}
 \bar{4}2m_{100} & 1 & 2_{001} & 2_{1\bar{1}0} & 2_{110} & \text{m}_{100} & \text{m}_{010} & \bar{4}^+_{001} & \bar{4}^-_{001} \\\hline
 \text{A}_1 & 1 & 1 & 1 & 1 & 1 & 1 & 1 & 1 \\
 \text{B}_1 & 1 & 1 & 1 & 1 & -1 & -1 & -1 & -1 \\
 \text{B}_2 & 1 & 1 & -1 & -1 & 1 & 1 & -1 & -1 \\
 \text{A}_2 & 1 & 1 & -1 & -1 & -1 & -1 & 1 & 1 \\
\end{array}
\end{align*}
\begin{align*}
\begin{array}{c|cccccccc}
 \text{4/m} & 1 & 2_{001} & 4^+_{001} & 4^-_{001} & \bar{1} & \text{m}_{001} & \bar{4}^+_{001} & \bar{4}^-_{001} \\\hline
 \text{A}_{\text{g}} & 1 & 1 & 1 & 1 & 1 & 1 & 1 & 1 \\
 \text{A}_{\text{u}} & 1 & 1 & 1 & 1 & -1 & -1 & -1 & -1 \\
 \text{B}_{\text{g}} & 1 & 1 & -1 & -1 & 1 & 1 & -1 & -1 \\
 \text{B}_{\text{u}} & 1 & 1 & -1 & -1 & -1 & -1 & 1 & 1 \\
 {}^2\text{E}_{\text{g}} & 1 & -1 & i & -i & 1 & -1 & i & -i \\
 {}^2\text{E}_{\text{u}} & 1 & -1 & i & -i & -1 & 1 & -i & i \\
 {}^1\text{E}_{\text{g}} & 1 & -1 & -i & i & 1 & -1 & -i & i \\
 {}^1\text{E}_{\text{u}} & 1 & -1 & -i & i & -1 & 1 & i & -i \\
\end{array}
&&
\begin{array}{c|cccccccc}
 \text{4mm} & 1 & 2_{001} & 4^+_{001} & 4^-_{001} & \text{m}_{100} & \text{m}_{010} & \text{m}_{1\bar{1}0} & \text{m}_{110} \\\hline
 \text{A}_1 & 1 & 1 & 1 & 1 & 1 & 1 & 1 & 1 \\
 \text{A}_2 & 1 & 1 & 1 & 1 & -1 & -1 & -1 & -1 \\
 \text{B}_1 & 1 & 1 & -1 & -1 & 1 & 1 & -1 & -1 \\
 \text{B}_2 & 1 & 1 & -1 & -1 & -1 & -1 & 1 & 1 \\
\end{array}
\end{align*}
\begin{align*}
\begin{array}{c|cccccccccccc}
 23 & 1 & 2_{100} & 2_{010} & 2_{001} & 3^-_{111} & 3^-_{\bar{1}\bar{1}1} & 3^-_{1\bar{1}\bar{1}} & 3^-_{\bar{1}1\bar{1}} & 3^+_{111} & 3^+_{\bar{1}1\bar{1}} & 3^+_{\bar{1}\bar{1}1} & 3^+_{1\bar{1}\bar{1}} \\\hline
 \text{A} & 1 & 1 & 1 & 1 & 1 & 1 & 1 & 1 & 1 & 1 & 1 & 1 \\
 {}^2\text{E} & 1 & 1 & 1 & 1 & e^{\frac{2 \pi i}{3}} & e^{\frac{2 \pi i}{3}} & e^{\frac{2 \pi i}{3}} & e^{\frac{2 \pi i}{3}} & e^{-\frac{2 \pi i}{3}} & e^{-\frac{2 \pi i}{3}} & e^{-\frac{2 \pi i}{3}} & e^{-\frac{2 \pi i}{3}} \\
 {}^1\text{E} & 1 & 1 & 1 & 1 & e^{-\frac{2 \pi i}{3}} & e^{-\frac{2 \pi i}{3}} & e^{-\frac{2 \pi i}{3}} & e^{-\frac{2 \pi i}{3}} & e^{\frac{2 \pi i}{3}} & e^{\frac{2 \pi i}{3}} & e^{\frac{2 \pi i}{3}} & e^{\frac{2 \pi i}{3}} \\
\end{array}
\end{align*}
\begin{align*}
\begin{array}{c|cccccccccccc}
 622 & 1 & 2_{100} & 2_{120} & 2_{001} & 6^+_{001} & 3^+_{001} & 3^-_{001} & 6^-_{001} & 2_{210} & 2_{110} & 2_{010} & 2_{1\bar{1}0} \\\hline
 \text{A}_1 & 1 & 1 & 1 & 1 & 1 & 1 & 1 & 1 & 1 & 1 & 1 & 1 \\
 \text{B}_2 & 1 & 1 & -1 & -1 & -1 & 1 & 1 & -1 & -1 & 1 & 1 & -1 \\
 \text{B}_1 & 1 & -1 & 1 & -1 & -1 & 1 & 1 & -1 & 1 & -1 & -1 & 1 \\
 \text{A}_2 & 1 & -1 & -1 & 1 & 1 & 1 & 1 & 1 & -1 & -1 & -1 & -1 \\
\end{array}
\end{align*}
\begin{align*}
\begin{array}{c|cccccccccccc}
 \bar{3}m_{100} & 1 & 2_{100} & \bar{1} & \text{m}_{100} & 3^+_{001} & 3^-_{001} & 2_{110} & 2_{010} & \bar{3}^+_{001} & \bar{3}^-_{001} & \text{m}_{110} & \text{m}_{010} \\\hline
 \text{A}_{\text{1g}} & 1 & 1 & 1 & 1 & 1 & 1 & 1 & 1 & 1 & 1 & 1 & 1 \\
 \text{A}_{\text{1u}} & 1 & 1 & -1 & -1 & 1 & 1 & 1 & 1 & -1 & -1 & -1 & -1 \\
 \text{A}_{\text{2g}} & 1 & -1 & 1 & -1 & 1 & 1 & -1 & -1 & 1 & 1 & -1 & -1 \\
 \text{A}_{\text{2u}} & 1 & -1 & -1 & 1 & 1 & 1 & -1 & -1 & -1 & -1 & 1 & 1 \\
\end{array}
\end{align*}
\begin{align*}
\begin{array}{c|cccccccccccc}
 \bar{6}2_{100}m_{210} & 1 & 2_{100} & \text{m}_{120} & \text{m}_{001} & 3^+_{001} & 3^-_{001} & 2_{110} & 2_{010} & \bar{6}^+_{001} & \bar{6}^-_{001} & \text{m}_{210} & \text{m}_{1\bar{1}0} \\\hline
 \text{A'}_1 & 1 & 1 & 1 & 1 & 1 & 1 & 1 & 1 & 1 & 1 & 1 & 1 \\
 \text{A''}_1 & 1 & 1 & -1 & -1 & 1 & 1 & 1 & 1 & -1 & -1 & -1 & -1 \\
 \text{A''}_2 & 1 & -1 & 1 & -1 & 1 & 1 & -1 & -1 & -1 & -1 & 1 & 1 \\
 \text{A'}_2 & 1 & -1 & -1 & 1 & 1 & 1 & -1 & -1 & 1 & 1 & -1 & -1 \\
\end{array}
\end{align*}
\begin{align*}
\begin{array}{c|cccccccccccc}
 \bar{3}\text{m}_{210} & 1 & 2_{120} & \bar{1} & \text{m}_{120} & 3^+_{001} & 3^-_{001} & 2_{210} & 2_{1\bar{1}0} & \bar{3}^+_{001} & \bar{3}^-_{001} & \text{m}_{210} & \text{m}_{1\bar{1}0} \\\hline
 \text{A}_{\text{1g}} & 1 & 1 & 1 & 1 & 1 & 1 & 1 & 1 & 1 & 1 & 1 & 1 \\
 \text{A}_{\text{1u}} & 1 & 1 & -1 & -1 & 1 & 1 & 1 & 1 & -1 & -1 & -1 & -1 \\
 \text{A}_{\text{2g}} & 1 & -1 & 1 & -1 & 1 & 1 & -1 & -1 & 1 & 1 & -1 & -1 \\
 \text{A}_{\text{2u}} & 1 & -1 & -1 & 1 & 1 & 1 & -1 & -1 & -1 & -1 & 1 & 1 \\
\end{array}
\end{align*}
\end{table*}

\begin{table*}
\caption*{(Continued from the previous page)}
\fontsize{7pt}{9pt}\selectfont
\begin{align*}
\begin{array}{c|cccccccccccc}
 \bar{6}2_{210}\text{m}_{100} & 1 & 2_{120} & \text{m}_{100} & \text{m}_{001} & 3^+_{001} & 3^-_{001} & 2_{210} & 2_{1\bar{1}0} & \bar{6}^+_{001} & \bar{6}^-_{001} & \text{m}_{110} & \text{m}_{010} \\\hline
 \text{A'}_1 & 1 & 1 & 1 & 1 & 1 & 1 & 1 & 1 & 1 & 1 & 1 & 1 \\
 \text{A''}_1 & 1 & 1 & -1 & -1 & 1 & 1 & 1 & 1 & -1 & -1 & -1 & -1 \\
 \text{A''}_2 & 1 & -1 & 1 & -1 & 1 & 1 & -1 & -1 & -1 & -1 & 1 & 1 \\
 \text{A'}_2 & 1 & -1 & -1 & 1 & 1 & 1 & -1 & -1 & 1 & 1 & -1 & -1 \\
\end{array}
\end{align*}
\arraycolsep=0.5pt
\begin{align*}
\begin{array}{c|cccccccccccccccccccccccc}
 432 & 1 & 2_{100} & 2_{010} & 2_{001} & 3^-_{111} & 3^-_{\bar{1}\bar{1}1} & 3^-_{1\bar{1}\bar{1}} & 3^-_{\bar{1}1\bar{1}} & 3^+_{111} & 3^+_{\bar{1}1\bar{1}} & 3^+_{\bar{1}\bar{1}1} & 3^+_{1\bar{1}\bar{1}} & 2_{1\bar{1}0} & 4^+_{001} & 4^-_{001} & 2_{110} & 2_{01\bar{1}} & 2_{011} & 4^+_{100} & 4^-_{100} & 2_{\bar{1}01} & 4^-_{010} & 2_{101} & 4^+_{010} \\\hline
 \text{A}_1 & 1 & 1 & 1 & 1 & 1 & 1 & 1 & 1 & 1 & 1 & 1 & 1 & 1 & 1 & 1 & 1 & 1 & 1 & 1 & 1 & 1 & 1 & 1 & 1 \\
 \text{A}_2 & 1 & 1 & 1 & 1 & 1 & 1 & 1 & 1 & 1 & 1 & 1 & 1 & -1 & -1 & -1 & -1 & -1 & -1 & -1 & -1 & -1 & -1 & -1 & -1 \\
\end{array}
\end{align*}
\begin{align*}
\begin{array}{c|cccccccccccc}
 \text{6/m} & 1 & 2_{001} & \bar{1} & \text{m}_{001} & 6^+_{001} & 3^+_{001} & 3^-_{001} & 6^-_{001} & \bar{6}^+_{001} & \bar{3}^+_{001} & \bar{3}^-_{001} & \bar{6}^-_{001} \\\hline
 \text{A}_{\text{g}} & 1 & 1 & 1 & 1 & 1 & 1 & 1 & 1 & 1 & 1 & 1 & 1 \\
 {}^1\text{E}_{\text{2g}} & 1 & 1 & 1 & 1 & e^{\frac{2 \pi i}{3}} & e^{-\frac{2 \pi i}{3}} & e^{\frac{2 \pi i}{3}} & e^{-\frac{2 \pi i}{3}} & e^{\frac{2 \pi i}{3}} & e^{-\frac{2 \pi i}{3}} & e^{\frac{2 \pi i}{3}} & e^{-\frac{2 \pi i}{3}} \\
 {}^2\text{E}_{\text{2g}} & 1 & 1 & 1 & 1 & e^{-\frac{2 \pi i}{3}} & e^{\frac{2 \pi i}{3}} & e^{-\frac{2 \pi i}{3}} & e^{\frac{2 \pi i}{3}} & e^{-\frac{2 \pi i}{3}} & e^{\frac{2 \pi i}{3}} & e^{-\frac{2 \pi i}{3}} & e^{\frac{2 \pi i}{3}} \\
 \text{A}_{\text{u}} & 1 & 1 & -1 & -1 & 1 & 1 & 1 & 1 & -1 & -1 & -1 & -1 \\
 {}^1\text{E}_{\text{2u}} & 1 & 1 & -1 & -1 & e^{\frac{2 \pi i}{3}} & e^{-\frac{2 \pi i}{3}} & e^{\frac{2 \pi i}{3}} & e^{-\frac{2 \pi i}{3}} & e^{-\frac{\pi i}{3}} & e^{\frac{\pi i}{3}} & e^{-\frac{\pi i}{3}} & e^{\frac{\pi i}{3}} \\
 {}^2\text{E}_{\text{2u}} & 1 & 1 & -1 & -1 & e^{-\frac{2 \pi i}{3}} & e^{\frac{2 \pi i}{3}} & e^{-\frac{2 \pi i}{3}} & e^{\frac{2 \pi i}{3}} & e^{\frac{\pi i}{3}} & e^{-\frac{\pi i}{3}} & e^{\frac{\pi i}{3}} & e^{-\frac{\pi i}{3}} \\
 {}^2\text{E}_{\text{1g}} & 1 & -1 & 1 & -1 & e^{\frac{\pi i}{3}} & e^{\frac{2 \pi i}{3}} & e^{-\frac{2 \pi i}{3}} & e^{-\frac{\pi i}{3}} & e^{\frac{\pi i}{3}} & e^{\frac{2 \pi i}{3}} & e^{-\frac{2 \pi i}{3}} & e^{-\frac{\pi i}{3}} \\
\end{array}
\end{align*}
\begin{align*}
\begin{array}{c|cccccccccccc}
 \text{6mm} & 1 & 2_{001} & \text{m}_{100} & \text{m}_{120} & 6^+_{001} & 3^+_{001} & 3^-_{001} & 6^-_{001} & \text{m}_{210} & \text{m}_{110} & \text{m}_{010} & \text{m}_{1\bar{1}0} \\\hline
 \text{A}_1 & 1 & 1 & 1 & 1 & 1 & 1 & 1 & 1 & 1 & 1 & 1 & 1 \\
 \text{A}_2 & 1 & 1 & -1 & -1 & 1 & 1 & 1 & 1 & -1 & -1 & -1 & -1 \\
 \text{B}_1 & 1 & -1 & 1 & -1 & -1 & 1 & 1 & -1 & -1 & 1 & 1 & -1 \\
 \text{B}_2 & 1 & -1 & -1 & 1 & -1 & 1 & 1 & -1 & 1 & -1 & -1 & 1 \\
\end{array}
\end{align*}
\begin{align*}
\begin{array}{c|cccccccccccccccc}
 \text{4/mmm} & 1 & 2_{100} & 2_{010} & 2_{001} & 2_{1\bar{1}0} & 4^+_{001} & 4^-_{001} & 2_{110} & \bar{1} & \text{m}_{100} & \text{m}_{010} & \text{m}_{001} & \text{m}_{1\bar{1}0} & \bar{4}^+_{001} & \bar{4}^-_{001} & \text{m}_{110} \\\hline
 \text{A}_{\text{1g}} & 1 & 1 & 1 & 1 & 1 & 1 & 1 & 1 & 1 & 1 & 1 & 1 & 1 & 1 & 1 & 1 \\
 \text{A}_{\text{1u}} & 1 & 1 & 1 & 1 & 1 & 1 & 1 & 1 & -1 & -1 & -1 & -1 & -1 & -1 & -1 & -1 \\
 \text{B}_{\text{1g}} & 1 & 1 & 1 & 1 & -1 & -1 & -1 & -1 & 1 & 1 & 1 & 1 & -1 & -1 & -1 & -1 \\
 \text{B}_{\text{1u}} & 1 & 1 & 1 & 1 & -1 & -1 & -1 & -1 & -1 & -1 & -1 & -1 & 1 & 1 & 1 & 1 \\
 \text{B}_{\text{2g}} & 1 & -1 & -1 & 1 & 1 & -1 & -1 & 1 & 1 & -1 & -1 & 1 & 1 & -1 & -1 & 1 \\
 \text{B}_{\text{2u}} & 1 & -1 & -1 & 1 & 1 & -1 & -1 & 1 & -1 & 1 & 1 & -1 & -1 & 1 & 1 & -1 \\
 \text{A}_{\text{2g}} & 1 & -1 & -1 & 1 & -1 & 1 & 1 & -1 & 1 & -1 & -1 & 1 & -1 & 1 & 1 & -1 \\
 \text{A}_{\text{2u}} & 1 & -1 & -1 & 1 & -1 & 1 & 1 & -1 & -1 & 1 & 1 & -1 & 1 & -1 & -1 & 1 \\
\end{array}
\end{align*}
\arraycolsep=0.5pt
\begin{align*}
\begin{array}{c|cccccccccccccccccccccccc}
 \text{m}\bar{3} & 1 & 2_{100} & 2_{010} & 2_{001} & 3^-_{111} & 3^-_{\bar{1}\bar{1}1} & 3^-_{1\bar{1}\bar{1}} & 3^-_{\bar{1}1\bar{1}} & 3^+_{111} & 3^+_{\bar{1}1\bar{1}} & 3^+_{\bar{1}\bar{1}1} & 3^+_{1\bar{1}\bar{1}} & \bar{1} & \text{m}_{100} & \text{m}_{010} & \text{m}_{001} & \bar{3}^-_{111} & \bar{3}^-_{\bar{1}\bar{1}1} & \bar{3}^-_{1\bar{1}\bar{1}} & \bar{3}^-_{\bar{1}1\bar{1}} & \bar{3}^+_{111} & \bar{3}^+_{\bar{1}1\bar{1}} & \bar{3}^+_{\bar{1}\bar{1}1} & \bar{3}^+_{1\bar{1}\bar{1}} \\\hline
 \text{A}_{\text{g}} & 1 & 1 & 1 & 1 & 1 & 1 & 1 & 1 & 1 & 1 & 1 & 1 & 1 & 1 & 1 & 1 & 1 & 1 & 1 & 1 & 1 & 1 & 1 & 1 \\
 \text{A}_{\text{u}} & 1 & 1 & 1 & 1 & 1 & 1 & 1 & 1 & 1 & 1 & 1 & 1 & -1 & -1 & -1 & -1 & -1 & -1 & -1 & -1 & -1 & -1 & -1 & -1 \\
 {}^2\text{E}_{\text{g}} & 1 & 1 & 1 & 1 & e^{\frac{2 \pi i}{3}} & e^{\frac{2 \pi i}{3}} & e^{\frac{2 \pi i}{3}} & e^{\frac{2 \pi i}{3}} & e^{-\frac{2 \pi i}{3}} & e^{-\frac{2 \pi i}{3}} & e^{-\frac{2 \pi i}{3}} & e^{-\frac{2 \pi i}{3}} & 1 & 1 & 1 & 1 & e^{\frac{2 \pi i}{3}} & e^{\frac{2 \pi i}{3}} & e^{\frac{2 \pi i}{3}} & e^{\frac{2 \pi i}{3}} & e^{-\frac{2 \pi i}{3}} & e^{-\frac{2 \pi i}{3}} & e^{-\frac{2 \pi i}{3}} & e^{-\frac{2 \pi i}{3}} \\
 {}^2\text{E}_{\text{u}} & 1 & 1 & 1 & 1 & e^{\frac{2 \pi i}{3}} & e^{\frac{2 \pi i}{3}} & e^{\frac{2 \pi i}{3}} & e^{\frac{2 \pi i}{3}} & e^{-\frac{2 \pi i}{3}} & e^{-\frac{2 \pi i}{3}} & e^{-\frac{2 \pi i}{3}} & e^{-\frac{2 \pi i}{3}} & -1 & -1 & -1 & -1 & e^{-\frac{\pi i}{3}} & e^{-\frac{\pi i}{3}} & e^{-\frac{\pi i}{3}} & e^{-\frac{\pi i}{3}} & e^{\frac{\pi i}{3}} & e^{\frac{\pi i}{3}} & e^{\frac{\pi i}{3}} & e^{\frac{\pi i}{3}} \\
\end{array}
\end{align*}
\begin{align*}
\begin{array}{c|cccccccccccccccccccccccc}
 \bar{4}3\text{m} & 1 & 2_{100} & 2_{010} & 2_{001} & 3^-_{111} & 3^-_{\bar{1}\bar{1}1} & 3^-_{1\bar{1}\bar{1}} & 3^-_{\bar{1}1\bar{1}} & 3^+_{111} & 3^+_{\bar{1}1\bar{1}} & 3^+_{\bar{1}\bar{1}1} & 3^+_{1\bar{1}\bar{1}} & \text{m}_{1\bar{1}0} & \bar{4}^+_{001} & \bar{4}^-_{001} & \text{m}_{110} & \text{m}_{01\bar{1}} & \text{m}_{011} & \bar{4}^+_{100} & \bar{4}^-_{100} & \text{m}_{\bar{1}01} & \bar{4}^-_{010} & \text{m}_{101} & \bar{4}^+_{010} \\\hline
 \text{A}_1 & 1 & 1 & 1 & 1 & 1 & 1 & 1 & 1 & 1 & 1 & 1 & 1 & 1 & 1 & 1 & 1 & 1 & 1 & 1 & 1 & 1 & 1 & 1 & 1 \\
 \text{A}_2 & 1 & 1 & 1 & 1 & 1 & 1 & 1 & 1 & 1 & 1 & 1 & 1 & -1 & -1 & -1 & -1 & -1 & -1 & -1 & -1 & -1 & -1 & -1 & -1 \\
\end{array}
\end{align*}
\end{table*}

\begin{table*}
\caption*{(Continued from the previous page)}
\fontsize{7pt}{9pt}\selectfont
\arraycolsep=0.5pt

\begin{align*}
%
 \right\} & 
 \\ \\\hline
 \text{A}_{\text{1g}} &1&1&1&1&1&1&1&1&1&1\\
 \text{A}_{\text{1u}} &1&1&1&1&1&-1&-1&-1&-1&-1\\
 \text{A}_{\text{2g}} &1&1&1&-1&-1&1&1&1&-1&-1\\
 \text{A}_{\text{2u}} &1&1&1&-1&-1&-1&-1&-1&1&1\\
\end{array}
\end{align*}
\end{table*}

\bibliography{refs}

\begin{thebibliography}{10}

\bibitem{HasanKaneRMP}
M.~Z. Hasan and C.~L. Kane.
\newblock Colloquium.
\newblock {\em Rev. Mod. Phys.}, 82:3045--3067, Nov 2010.

\bibitem{QiZhangRMP}
Xiao-Liang Qi and Shou-Cheng Zhang.
\newblock Topological insulators and superconductors.
\newblock {\em Rev. Mod. Phys.}, 83:1057--1110, Oct 2011.

\bibitem{RyuTenFold}
Shinsei Ryu, Andreas~P. Schnyder, Akira Furusaki, and Andreas~W.W. Ludwig.
\newblock Topological insulators and superconductors: tenfold way and
  dimensional hierarchy.
\newblock {\em New Journal of Physics}, 12(6):065010, 2010.

\bibitem{KitaevPeriodic}
Alexei Kitaev.
\newblock Periodic table for topological insulators and superconductors.
\newblock {\em AIP Conference Proceedings}, 1134(1):22--30, 2009.

\bibitem{Stone2010}
Michael Stone, Ching-Kai Chiu, and Abhishek Roy.
\newblock Symmetries, dimensions and topological insulators: the mechanism
  behind the face of the bott clock.
\newblock {\em Journal of Physics A: Mathematical and Theoretical},
  44(4):045001, dec 2010.

\bibitem{WenFreeFermion}
Xiao-Gang Wen.
\newblock Symmetry-protected topological phases in noninteracting fermion
  systems.
\newblock {\em Phys. Rev. B}, 85:085103, Feb 2012.

\bibitem{MorimotoClifford}
Takahiro Morimoto and Akira Furusaki.
\newblock Topological classification with additional symmetries from clifford
  algebras.
\newblock {\em Phys. Rev. B}, 88:125129, Sep 2013.

\bibitem{HermeleBuilding}
Sheng-Jie Huang, Hao Song, Yi-Ping Huang, and Michael Hermele.
\newblock Building crystalline topological phases from lower-dimensional
  states.
\newblock {\em Phys. Rev. B}, 96:205106, Nov 2017.

\bibitem{LukaHigherOrder}
Luka Trifunovic and Piet~W. Brouwer.
\newblock Higher-order bulk-boundary correspondence for topological crystalline
  phases.
\newblock {\em Phys. Rev. X}, 9:011012, Jan 2019.

\bibitem{KSGeneHomo}
Ken Shiozaki, Charles~Zhaoxi Xiong, and Kiyonori Gomi.
\newblock Generalized homology and atiyah-hirzebruch spectral sequence in
  crystalline symmetry protected topological phenomena.
\newblock arXiv:1810.00801.

\bibitem{BenalcazarScience}
Wladimir~A. Benalcazar, B.~Andrei Bernevig, and Taylor~L. Hughes.
\newblock Quantized electric multipole insulators.
\newblock {\em Science}, 357(6346):61--66, 2017.

\bibitem{FangFu17}
Chen Fang and Liang Fu.
\newblock New classes of topological crystalline insulators having surface
  rotation anomaly.
\newblock {\em Science Advances}, 5(12), 2019.

\bibitem{SchindlerHigher}
Frank Schindler, Ashley~M. Cook, Maia~G. Vergniory, Zhijun Wang, Stuart S.~P.
  Parkin, B.~Andrei Bernevig, and Titus Neupert.
\newblock Higher-order topological insulators.
\newblock {\em Science Advances}, 4(6), 2018.

\bibitem{KhalafPRX18AII}
Eslam Khalaf, Hoi~Chun Po, Ashvin Vishwanath, and Haruki Watanabe.
\newblock Symmetry indicators and anomalous surface states of topological
  crystalline insulators.
\newblock {\em Phys. Rev. X}, 8:031070, Sep 2018.

\bibitem{SongMapping}
Zhida Song, Tiantian Zhang, Zhong Fang, and Chen Fang.
\newblock Quantitative mappings between symmetry and topology in solids.
\newblock {\em Nature communications}, 9(1):3530, 2018.

\bibitem{HermeleTopologicalCrystal}
Zhida Song, Sheng-Jie Huang, Yang Qi, Chen Fang, and Michael Hermele.
\newblock Topological states from topological crystals.
\newblock {\em Science Advances}, 5(12), 2019.

\bibitem{Haruki230}
Hoi~Chun Po, Ashvin Vishwanath, and Haruki Watanabe.
\newblock Complete theory of symmetry-based indicators of band topology.
\newblock {\em Nat. Commun.}, 8(1):50, 2017.

\bibitem{CornfeldPointGroup}
Eyal Cornfeld and Adam Chapman.
\newblock Classification of crystalline topological insulators and
  superconductors with point group symmetries.
\newblock {\em Phys. Rev. B}, 99:075105, Feb 2019.

\bibitem{AZ}
Alexander Altland and Martin~R. Zirnbauer.
\newblock Nonstandard symmetry classes in mesoscopic normal-superconducting
  hybrid structures.
\newblock {\em Phys. Rev. B}, 55:1142--1161, Jan 1997.

\bibitem{Teo-Kane}
Jeffrey C.~Y. Teo and C.~L. Kane.
\newblock Topological defects and gapless modes in insulators and
  superconductors.
\newblock {\em Phys. Rev. B}, 82:115120, Sep 2010.

\bibitem{KS_Atiyah}
Ken Shiozaki, Masatoshi Sato, and Kiyonori Gomi.
\newblock Atiyah-hirzebruch spectral sequence in band topology: General
  formalism and topological invariants for 230 space groups.
\newblock arXiv:1802.06694.

\bibitem{OkumaSatoShiozaki}
Nobuyuki Okuma, Masatoshi Sato, and Ken Shiozaki.
\newblock Topological classification under nonmagnetic and magnetic point group
  symmetry: Application of real-space atiyah-hirzebruch spectral sequence to
  higher-order topology.
\newblock {\em Phys. Rev. B}, 99:085127, Feb 2019.

\end{thebibliography}
\bibliographystyle{unsrt} 
\end{document}